\documentclass[aip,reprint]{revtex4-1}

\usepackage{bm}
\usepackage{natbib}
\usepackage{graphicx}
\usepackage{color}
\usepackage{hyperref}

\hypersetup{
    colorlinks,%
    citecolor=blue,%
    linkcolor=blue,%
    urlcolor=blue
}

\begin{document}


\title{Gamma-ray flares in the Crab Nebula: A case of relativistic reconnection?} 



\author{B.~Cerutti}
\email[bcerutti@astro.princeton.edu]
\altaffiliation{Lyman Spitzer Jr. Fellow}
\affiliation{Department of Astrophysical Sciences, Princeton University, Princeton, NJ 08544, USA}

\author{G.~R.~Werner}
\email[greg.werner@colorado.edu]
\affiliation{Center for Integrated Plasma Studies, Physics Department, University of Colorado, UCB 390, Boulder, CO 80309-0390, USA}

\author{D.~A.~Uzdensky}
\email[uzdensky@colorado.edu]
\affiliation{Center for Integrated Plasma Studies, Physics Department, University of Colorado, UCB 390, Boulder, CO 80309-0390, USA}

\author{M.~C.~Begelman}
\email[mitch@jila.colorado.edu]
\altaffiliation{Department of Astrophysical and Planetary Sciences, University of Colorado, UCB 391, Boulder, CO 80309-0391, USA}
\affiliation{JILA, University of Colorado and National Institute of Standards and Technology, UCB 440, Boulder, CO 80309-0440, USA}


\date{\today}

\begin{abstract}
The Crab Nebula was formed after the collapse of a massive star about a thousand years ago, leaving behind a pulsar that inflates a bubble of ultra-relativistic electron-positron pairs permeated with magnetic field. The observation of brief but bright flares of energetic gamma rays suggests that pairs are accelerated to PeV energies within a few days; such rapid acceleration cannot be driven by shocks. Here, it is argued that the flares may be the smoking gun of magnetic dissipation in the Nebula. Using 2D and 3D particle-in-cell simulations, it is shown that the observations are consistent with relativistic magnetic reconnection, where pairs are subject to strong radiative cooling. The Crab flares may highlight the importance of relativistic magnetic reconnection in astrophysical sources.
\end{abstract}


\maketitle 


\section{Introduction}\label{introduction}

\begin{figure}
\includegraphics[scale=0.32]{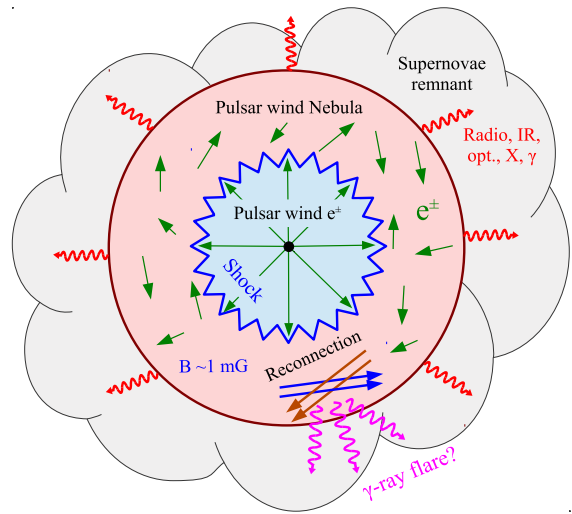}
\caption{This diagram sketches the classical (1D) model of pulsar wind nebulae in which the pulsar inflates a relativistic and magnetized wind of $e^{\pm}$ pairs. The pulsar wind nebula forms between the termination shock of the wind and the contact discontinuity with the ambient medium, here the material from the supernova remnant. As argued in this paper, magnetic reconnection within the nebula could explain the gamma-ray flares observed in the Crab Nebula. \label{fig1}}
\end{figure}

The Crab Nebula was born soon after the collapse of a massive star followed by a supernova explosion observed on Earth in 1054~AD. In classical models of pulsar wind nebulae\cite{1974MNRAS.167....1R, 1984ApJ...283..694K, 2009ASSL..357..421K, 2012SSRv..173..341A, 2013arXiv1309.7046B}, a rapidly rotating, highly magnetized neutron star, i.e., a pulsar, is constantly injecting energy into the nebula in the form of a magnetized, cold, and ultra-relativistic wind of electron and positron pairs. The pulsar wind nebula forms where the wind terminates its free expansion, at about $0.1$~pc in the Crab, and is confined by the material of the supernova remnant (Figure~\ref{fig1}). In the Crab Nebula, the pairs are randomized and radiate synchrotron radiation from radio to $100~$MeV gamma-rays and inverse Compton emission from $1~$GeV to about $100~$TeV by upscattering the low-energy background light \cite{1996MNRAS.278..525A, 2010A&A...523A...2M}.

In September 2010, the gamma-ray space telescopes {\em Agile} and {\em Fermi}-LAT reported the first detections of very bright flares of high-energy gamma rays ($>100$~MeV) from the Crab Nebula \cite{2011Sci...331..736T, 2011Sci...331..739A, 2011A&A...527L...4B}. This was a huge surprise for both teams because the nebula was a well-known constant emitter of gamma rays, so constant in fact that this source is used for instrumental calibration purposes. Since the discovery of the first flares, one flare every year or so appears in the gamma-ray sky: April 2011, July 2012, March 2013, and October 2013 \cite{2011ApJ...741L...5S, 2012ApJ...749...26B, 2013ApJ...765...52S, 2013ApJ...775L..37M, 2013arXiv1309.7046B, 2013ATel.5485....1B}. Outside of these remarkable episodes, the gamma-ray flux remains constantly variable \cite{2012ApJ...749...26B}, although with a much smaller amplitude, suggesting that the mechanism at the origin of the flares may be happening continuously in the nebula. 

A closer look at the flares reveals an even richer, but also more puzzling picture. The first striking feature of the flares is their duration: typically lasting for a few days to a couple of weeks. If one assumes that the flaring emitting region is causally connected, this implies an emitter of size $c t_{\rm flare}\sim 10^{16}~$cm, where $c$ is the velocity of light and $t_{\rm flare}=1~$week, which represents roughly $1\%$ of the size of the nebula that radiates $\sim 10$ times more flux than the entire system. During the brightest flares, one can even resolve intra-day variability timescales\cite{2011A&A...527L...4B, 2012ApJ...749...26B, 2013ApJ...775L..37M}. There is a consensus that the observed gamma-ray emission is synchrotron radiation (however, see Ref.~\onlinecite{2013ApJ...763..131T}), because this is the most efficient radiative process under the physical conditions found in the nebula. Hence, the particles at the origin of the flares are PeV ($10^{15}$~eV) electrons (and positrons) immersed in a milliGauss magnetic field, much stronger than the $\sim 100~\mu$G field usually expected\cite{2010A&A...523A...2M}. It turns out that such particles have a relativistic gyration time of order the duration of the flares themselves. In other words, the pairs must be accelerated and radiating over a sub-Larmor timescale. This causes serious challenges to classical models of particle acceleration, such as diffuse shock-acceleration, where the particles gain energy over several Larmor cycles. In addition, the gamma-ray spectrum and the non-detection of the flare at other wavelengths \cite{2013ApJ...765...56W} (in radio, infra-red, optical, X-rays) point towards a very narrow particle spectrum, perhaps even mono-energetic. This is another argument against particle acceleration via shock-acceleration.

The last challenging aspect of the flares, and arguably the most spectacular one, is the emission of synchrotron photons with an energy up to $400~$MeV, i.e., well above the standard $\epsilon_{\rm max}=160~$MeV synchrotron burn-off limit \cite{1983MNRAS.205..593G, 1996ApJ...457..253D}. This limit is set by the balance between the accelerating electric force acting on an electron or positron, $F_{\rm e}=eE$, where $e$ is the charge of the electron and $E$ the electric field, and the radiation reaction force induced by the emission of synchrotron radiation, $F_{\rm rad}\approx 2/3 r_{\rm e}^2\gamma^2 B_{\perp}^2$ (see Section~\ref{section_zeltron}), where $r_{\rm e}=2.82\times 10^{-13}~$cm is the classical radius of the electron, $\gamma$ is the Lorentz factor of the particle and $B_{\perp}$ is the magnetic field strength perpendicular to the particle's velocity vector. Setting $F_{\rm e}=F_{\rm rad}$ gives the maximum Lorentz factor reached by the particle, i.e.,
\begin{equation}
\gamma_{\rm rad}=\left(\frac{3 e E}{2 r_{\rm e}^2 B_{\perp}^2}\right)^{1/2}.
\label{grad}
\end{equation}
Using Eq.~(\ref{grad}), the critical synchrotron photon energy\cite{1970RvMP...42..237B} (i.e., where the synchrotron spectrum of a single particle peaks) emitted by the radiation-reaction-limited particles then equals
\begin{eqnarray}
\epsilon^{\rm max}_{\rm crit} & = &\frac{3 h e}{4\pi m_{\rm e}c}B_{\perp}\gamma_{\rm rad}^2 \nonumber \\
 & = & \frac{9}{4} \frac{m_{\rm e}c^2}{\alpha_{\rm F}}\left(\frac{E}{B_{\perp}}\right) \approx 160\left(\frac{E}{B_{\perp}}\right)~{\rm MeV},
\label{emax}
\end{eqnarray}
where $h$ is the Planck constant, $m_{\rm e}$ the mass of the electron, and $\alpha_{\rm F}\approx 1/137$ is the fine structure constant. Hence, the maximum synchrotron photon energy depends only on the ratio $E/B_{\perp}$. Under ideal magnetohydrodynamic (MHD) conditions, $E$ is smaller, or at most equal to $B_{\perp}$, hence the $160~$MeV limit.

To remain below this limit in the rest frame of the source, the flaring region must be moving towards the observer with a bulk velocity $>0.9 c$ to explain the $400~$MeV synchrotron photons in the Crab flares. Most models of the flares proposed so far rely on this assumption \cite{2011MNRAS.414.2229B, 2011MNRAS.414.2017K, 2011ApJ...730L..15Y, 2012MNRAS.422.3118L, 2012MNRAS.426.1374C, 2012MNRAS.427.1497L}. However, the measurements of proper motions in optical and in X-rays \cite{2002ApJ...577L..49H} show only mildly relativistic speeds in the nebula, up to $0.5 c$. The alternative point of view developed by the authors in a series of papers\cite{2011ApJ...737L..40U, 2012ApJ...746..148C, 2013ApJ...770..147C, 2013arXiv1311.2605C} supposes that particle acceleration does occur in a region where ideal MHD breaks down, i.e., where $E\gtrsim 2B_{\perp}$. This situation is naturally encountered in sites of magnetic reconnection \cite{2004PhRvL..92r1101K, 2007A&A...472..219C}, within the diffusion region where the magnetic field decreases and changes direction, and where the electric field is maximum. As reviewed below (Section~\ref{section_results}), the magnetic reconnection scenario is able to account for all the challenging aspects of the flares, and points towards a magnetized model of the Crab Nebula, i.e., with a low plasma-$\beta$, or with a high magnetization parameter $\sigma$ (ratio between the magnetic to particle enthalpy densities).

\section{Introducing Zeltron: a new PIC code with radiative feedback}\label{section_zeltron}

Particle-in-cell (PIC) simulations are well-suited for studying ultra-relativistic collisionless pair plasma reconnection. However, only a few codes\cite{2009PhRvL.103g5002J, 2010NJPh...12l3005T} take into account the effect of the radiation reaction force on the motion of the particles, an effect of critical importance in modeling the Crab flares (Section~\ref{introduction}). {\tt Zeltron} is a new PIC code that does have this ability, and was specifically developed from scratch for studying the Crab flare puzzle \cite{2013ApJ...770..147C}. Hence, instead of the Lorentz-Newton equation, the code solves the so-called ``Abraham-Lorentz-Dirac'' equation\cite{1975ctf..book.....L} for the motion of the particles. In a non-covariant form, this equation is given by
\begin{equation}
m_{\rm e}\frac{d\left(\gamma\mathbf{v}\right)}{dt}=\mathbf{F_{\rm L}}+\mathbf{g},
\label{eq_ald}
\end{equation}
where $\mathbf{v}=\bm{\beta_{\rm e}}c$ is the velocity of the particle, $\mathbf{F_{\rm L}}=\pm e(\mathbf{E}+\bm{\beta_{\rm e}}\times\mathbf{B})$ is the usual Lorentz force, and $\mathbf{g}$ is the synchrotron radiation reaction force. Computing $\mathbf{g}$ can be quite challenging in the general case, but in the ultra-relativistic limit ($\beta_{\rm e}\approx 1,~\gamma\gg1$) and in the classical electrodynamic regime, i.e., if $\gamma B/B_{\rm QED}\ll 1$, where $B_{\rm QED}=4.4\times 10^{13}~$G, the radiation reaction force has a remarkably simple formulation:
\begin{equation}
\mathbf{g}=-\frac{2}{3}r_{e}^2\gamma^2\left[\left(\mathbf{E}+\bm{\beta}\times\mathbf{B}\right)^2-\left(\bm{\beta}\cdot\mathbf{E}\right)^2\right]\bm{\beta},
\end{equation}
i.e., it is analogous to a friction force in classical mechanics. The implementation of the radiation reaction force in {\tt Zeltron} follows the modified version of the classical finite-difference time-domain (FDTD) Boris algorithm proposed by Tamburini et al. (2010) (Ref.~\onlinecite{2010NJPh...12l3005T}). The current and charge densities are then deduced at the grid points from the position and the velocity of the particles. Finally, Maxwell's equations are advanced in time using the FDTD Yee algorithm \cite{1966ITAP...14..302Y}. The code is efficiently parallelized in three dimensions with the message passing interface MPI using a domain decomposition technique.

\section{Simulating the flares}\label{section_results}

This section summarizes the main results of a series of 2D and 3D PIC simulations performed with {\tt Zeltron} of relativistic pair plasma reconnection with radiation reaction force in the Crab Nebula.

\subsection{2D and 3D PIC simulations}

The computational domain is a square of dimension $L_{\rm x}\times L_{\rm y}$ in 2D, and a cube of dimension $L_{\rm x}\times L_{\rm y} \times L_{\rm z}$ in 3D with periodic boundary conditions in all directions. All the simulations are initialized with the relativistic Harris equilibrium\cite{2003ApJ...591..366K}, i.e., a kinetic equilibrium where the gas pressure within the reconnection layer balances the upstream reconnecting magnetic pressure. To comply with the periodic boundary conditions, the box contains two anti-parallel layers of half-thickness $\delta$. The reconnecting magnetic field is along the $\pm x$-directions, $B_{\rm x}$, and reverses across each layer at $y=0.25 L_{\rm y}/4$ and $y=0.75 L_{\rm y}$. The reconnecting field far upstream, $B_0$, is set at $5$~mG, so that $\gamma_{\rm rad}\approx 1.3\times 10^9$ [for $E=B_{\perp}$ in Eq.~(\ref{grad})]. The guide field is along the $z$-direction and is a free parameter of the simulation.

The initial current flowing within each layer along the $z$-direction is carried by a population of ``drifting'' pairs moving at a mildly relativistic bulk velocity $v_{\rm drift}=0.6 c$. Another pair plasma is injected uniformly throughout the box with a density 10 times smaller than the drifting plasma density found in the layers ($n_{\rm bg}=0.1 n_0$). Both plasmas are injected with a relativistic Maxwellian distribution with a temperature $\theta_0\equiv kT_0/m_{\rm e}c^2=10^8$. The magnetization parameter defined in Section~\ref{introduction} is then $\sigma\equiv B_0^2/8\pi(0.1 n_0)\theta_0 m_{\rm e}c^2\approx 16$. For more details about the properties of the simulations, see Refs.~\onlinecite{2013ApJ...770..147C, 2013arXiv1311.2605C}.

\begin{figure}
\centering
\includegraphics[scale=0.35]{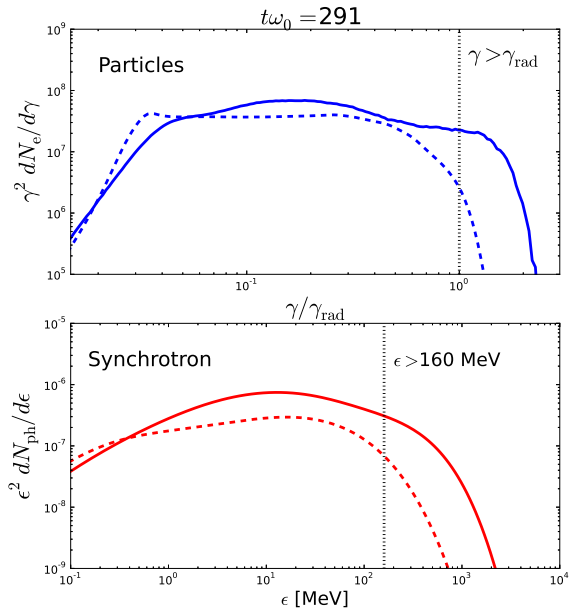}
\caption{Total particle ($\gamma^2 dN_{\rm e}/d\gamma$, top panel) and optically thin synchrotron radiation ($\epsilon^2 dN_{\rm ph}/d\epsilon$, bottom panel) energy distributions averaged over all directions, in a 2D simulation of size $L_{\rm x}=L_{\rm y}=200\rho_0$ without a guide field (solid lines), and in a 3D simulation of size $L_{\rm x}=L_{\rm y}=L_{\rm z}=200\rho_0$ with $B_{\rm z}=0.5 B_0$ (dashed lines)\cite{2013arXiv1311.2605C}. The vertical dotted lines represent the standard radiation reaction limited energy for the particles [top panel, Eq.~(\ref{grad})] and for the synchrotron photons [bottom panel, Eq.~(\ref{emax})]. \label{fig2}}
\end{figure}

In 2D, the thin reconnection layers quickly become unstable to tearing modes. As a result, the layers break up into chains of plasmoids separated by X-points where magnetic field lines reconnect. The strong magnetic tension released at X-points pushes the plasma along the $\pm x$-direction, and drives the merging of plasmoids into bigger ones. As shown in the next section (Section~\ref{speiser}), the particles are accelerated beyond $\gamma_{\rm rad}$ [Eq.~(\ref{grad})] by the reconnection electric field at X-points, where $E\gg B_{\perp}$. Figure~\ref{fig2} shows the total particle and optically thin synchrotron radiation energy distributions averaged over all directions in a 2D simulation (solid lines), with no guide field, of dimension $L_{\rm x}=L_{\rm y}=200\rho_0$, where $\rho_0=\theta_0 m_{\rm e}c^2/eB_0\approx 3.4\times 10^{13}~$cm, at the time $t=291\omega_0^{-1}$, with $\omega_0\equiv c/\rho_0\approx 8.8\times 10^{-4}~$s$^{-1}$. The maximum energy reached by the particles, $\gamma_{\rm max}\gtrsim 2\gamma_{\rm rad}$, as well as the maximum synchrotron photon energy, $\epsilon_{\rm max}\gtrsim 1~$GeV, are substantially above the radiation reaction limit.

In 3D, the layers are tearing and kink unstable. The tearing creates a network of magnetic flux ropes, similar to plasmoids in 2D, separated by X-lines. In the zero-guide field case, the kink instability deforms the layers along the $z$-direction and eventually leads to their disruption with particle heating instead of non-thermal particle acceleration. However, a moderate guide field, $B_{\rm z}\gtrsim 0.5 B_0$, is enough to suppress the deformation of the layer without changing significantly the growth rates of the fastest tearing modes \cite{2008ApJ...677..530Z, 2013arXiv1311.2605C}. In this case, non-thermal particle acceleration can operate along the X-lines, as in 2D, and the highest-energy particles are accelerated above the standard radiation reaction limit and emit $>160$~MeV synchrotron radiation (Figure~\ref{fig2}, dashed lines).

\subsection{Extreme particle acceleration mechanism}\label{speiser}

\begin{figure}
\centering
\includegraphics[scale=0.55]{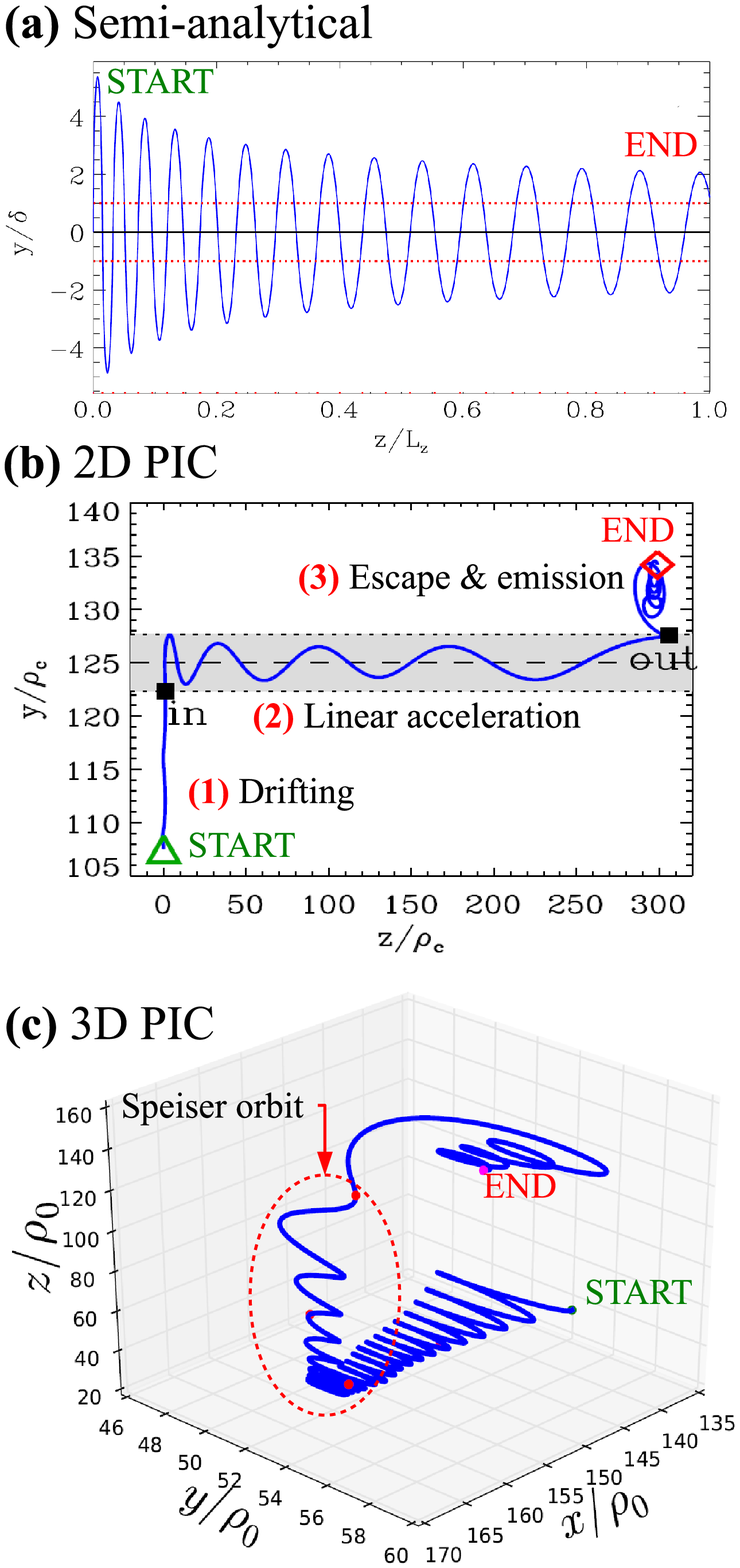}
\caption{Single particle trajectory accelerated above the standard radiation reaction limit energy [Eq.~(\ref{grad})] by the reconnection electric field along X-points/X-lines. The orbit in panel (a) was obtained using a test particle simulation with radiation reaction force and with prescribed static fields\cite{2012ApJ...746..148C}. Panel (b)/(c) show the orbit of a tracked macroparticle in a 2D\cite{2013ApJ...770..147C}/3D\cite{2013arXiv1311.2605C} PIC simulation with radiation reaction force. Panel (b) presents the three distinct phases of particle acceleration and emission highlighted in the text and in Ref.~\onlinecite{2013ApJ...770..147C}, namely: (1) the drifting phase, (2) the linear acceleration phase, and (3) the phase where the particle escapes and emits $>160~$MeV synchrotron radiation. \label{fig3}}
\end{figure}

{\tt Zeltron} can track a large sub-sample of macroparticles throughout the simulation, which enables one to follow a statistically significant number of the highest-energy particles in the simulation. The analysis of the high-energy particle orbits reveals a very repeatable and simple pattern in the acceleration process, both in 2D and 3D PIC simulations. First, almost all of the highest-energy particles are not inside the reconnection layer initially, but far upstream. As reconnection proceeds in the plasmoid-dominated regime, particles stay in the $xy$-plane and drift towards the X-points, in general with no energy gain [phase 1, Figure~\ref{fig3} panel (b)].

However, as soon as the particles reach the X-points/X-lines, they become trapped in the direction across the layer by the reversing magnetic field and follow the relativistic analog of a Speiser orbit\cite{1965JGR....70.4219S} [phase 2, Figure~\ref{fig3} panel (b)]. In this phase, the energy of the particles increases almost linearly with time because they experience the strong, quasi-uniform reconnection electric force along the $z$-direction. The distance between the particle and the layer mid-plane decreases rapidly with time, in agreement with (semi-)analytical expectations\cite{2011ApJ...737L..40U, 2012ApJ...746..148C, 2013ApJ...770..147C} [Figure~\ref{fig3}, panel (a)]. This feature of Speiser orbits is fundamental for particle acceleration above the standard $\gamma_{\rm rad}$ [Eq.~(\ref{grad})], because it enables the particles to reach regions of very low magnetic field with $E\gg B_{\perp}$, and hence little radiation reaction force.

At this stage the particles are extremely energetic, with $\gamma>\gamma_{\rm rad}$, but they are not radiating. To emit synchrotron photons above $160~$MeV, the particles must return into a region of strong magnetic field, where $E\ll B_{\perp}$. In the simulation, this happens when the particles reach the final big magnetic island [phase 3, Figure~\ref{fig3} panel (b)]. The particles are then losing almost all their energy within a fraction of a full Larmor gyration in the form of a short pulse of $>160~$MeV radiation. Particles are deflected away from the X-lines sooner if a guide field is present. Hence, a strong guide field ($B_{\rm z}\gtrsim B_0$) suppresses the extreme particle acceleration mechanism described here and the associated emission of super-energetic synchrotron photons.

\subsection{Anisotropy, bunching, and radiative signatures}

\begin{figure}
\centering
\includegraphics[scale=0.42]{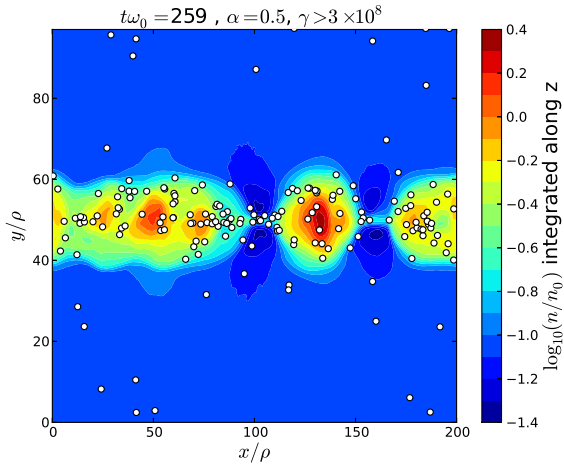}
\caption{Plasma density integrated along the $z$-direction in a 3D PIC simulations\cite{2013arXiv1311.2605C} with a $B_{\rm z}=0.5 B_0$ guide field. The white dots show the location of a sample of tracked high-energy particles with $\gamma>3\times10^8$. \label{fig4}}
\end{figure}

A direct consequence of particle acceleration via relativistic reconnection is spatial bunching of the energetic particles around X-points/X-lines (the acceleration regions) and within magnetic islands/flux ropes (the regions that collect the reconnection outflow and the energetic particles). Figure~\ref{fig4} shows the positions of a sample of tracked high-energy particles with $\gamma>3\times 10^8$ (white circles) in a 3D reconnection simulation of size $L_{\rm x}=L_{\rm y}=L_{\rm z}=200\rho_0$, in the presence of a moderate guide field $B_{\rm z}=0.5 B_0$, superimposed on the plasma density integrated along the $z$-direction (color-coded) towards the end of the simulation. Hence, reconnection transfers and concentrates the reconnecting magnetic energy initially distributed throughout the box into particles distributed over a much smaller volume. This result has important implications in astronomy\cite{2012ApJ...754L..33C, 2012MNRAS.425.2519N}, and in particular for the Crab flares\cite{2013ApJ...770..147C, 2013arXiv1311.2605C}, because one usually estimates the required energetics from the size of the emitting regions. Particle bunching allows one to relax the energetic constraints because the reservoir of free energy (here the box size, $L_{\rm x}$) can be much larger than the size of the emitting regions themselves (size of the plasmoids/flux ropes, here $\lesssim 0.1 L_{\rm x}$).

\begin{figure}
\centering
\includegraphics[scale=0.40]{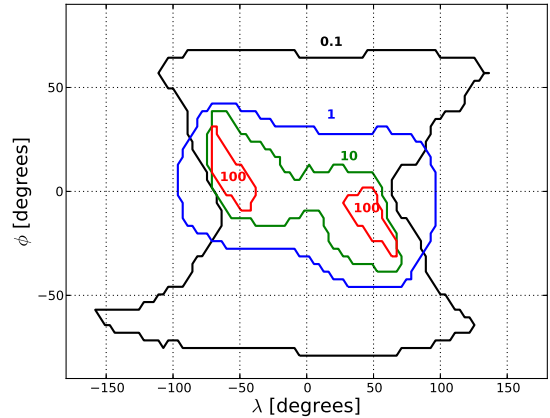}
\caption{Angular distribution of the synchrotron radiation inferred from a 3D PIC simulation\cite{2013arXiv1311.2605C} with a $B_{\rm z}=0.5 B_0$ guide field, at $t\omega_0^{-1}=291$. Each contour shows the angle spread where half of the photon flux is emitted in 4 photon energy bands: 0.1, 1, 10, and 100~MeV (from the outermost to the innermost contour). See the text for the definition of the angles $\lambda$ and $\phi$. \label{fig5}}
\end{figure}

Another crucial by-product of relativistic reconnection is the strong energy-dependent anisotropy of the particle velocities, the highest energy particles being the most focused. In addition, any anisotropy in the particle distribution translates into beaming of the emitted radiation, because relativistic particles emit photons along their direction of motion within a cone of semi-aperture angle $1/\gamma\ll 1$. Figure~\ref{fig5} presents the angular spread of the optically thin synchrotron radiation emitted throughout the box, in 4 photon energy bands: 0.1, 1, 10, and 100~MeV. The angular distributions are shown with respect to the angles $\phi$ and $\lambda$ such that a spherical radial unit vector is defined as $(\cos\phi\sin\lambda,\sin\phi,\cos\phi\cos\lambda)$ in the $(xyz)$-coordinate system. While the low energy radiation ($0.1$~MeV) is mostly isotropic, the highest energy radiation, of interest for the Crab flares ($>100~$MeV), is focused into two narrow beams concentrated at $\lambda\approx\pm 50$-$60^{\circ}$, which approximately corresponds to the direction of the upstream magnetic field line with a $0.5 B_0$ guide field. In the zero-guide field case, the beam points toward $\lambda\approx\pm 90^{\circ}$ (Refs.~\onlinecite{2012ApJ...754L..33C, 2013ApJ...770..147C}). In addition to bunching, the beaming of the $>100$~MeV synchrotron radiation diminishes the energy budget required to explain the apparent isotropic gamma-ray luminosity observed during the Crab flares. 

Each bunch of energetic particles produces a beam that moves rapidly within the reconnection layer\cite{}. Each time a beam crosses the observer's line of sight, there is an intense flux of $>160$~MeV synchrotron emission that lasts for about the light-crossing time of the bunch, corresponding to about $6$-$8$ hours in the simulations \cite{2013ApJ...770..147C, 2013arXiv1311.2605C}. This timescale is compatible with the shortest variability timescale measured by {\em Fermi} during the brightest gamma-ray flares in the Crab \cite{2012ApJ...749...26B, 2013ApJ...775L..37M}, while the overall duration of the flare is well explained by the global reconnection time, of order $L_{\rm x}/(\beta_{\rm rec}c)\sim 10$~days, where $\beta_{\rm rec}\sim 0.2$-$0.3$ is the dimensionless reconnection rate measured in the simulations. Simulated light curves (i.e., flux as a function of time) can be found in Refs.~\onlinecite{2013ApJ...770..147C, 2013arXiv1311.2605C}.

\section{Summary and implications}

The magnetic reconnection scenario addresses successfully several of the key features of the observed gamma-ray flares in the Crab Nebula. The most important result is the unambiguous discovery of particle acceleration above the standard radiation reaction limit, and the emission of $>160~$MeV synchrotron radiation in 2D and 3D PIC simulations of pair plasma reconnection with radiation reaction force. The acceleration mechanism is very robust and operates at X-points/X-lines where the particles follow relativistic Speiser orbits and are accelerated linearly by the reconnection electric field, where $E\gg B_{\perp}$.

A direct by-product of this overall acceleration process is the strong anisotropy of the energetic particles and radiation, the highest energy particles/radiation being the most beamed. This result has very important consequences for astronomers because the observed flux and the occurrence of each flare should depend on the orientation of the reconnection layer with respect to the observer. In addition, strong beaming helps to reduce the tight energetic constraints imposed by the flare, in particular for the highest energy radiation. Another implication of particle acceleration via reconnection is the spatial bunching of the energetic particles into the magnetic islands/flux ropes. Here again, this effect helps to alleviate the energetic constraints because the emitting regions (magnetic islands/flux ropes) can be much smaller than the system size where the reservoir of free energy, i.e., the magnetic energy, is initially present. The combination of particle bunching and beaming allows for rapid changes in time of the gamma-ray flux if the beam points towards Earth. Other features of the flares (not discussed here) can be well reproduced with this model as well, such as the spectral shape, the flux-energy correlation, and the power spectrum \cite{2013ApJ...770..147C}.

The reconnection scenario supposes that the nebula is, at least, mildly magnetized with $\sigma\gtrsim 1$, contrary to classical models\cite{1974MNRAS.167....1R, 1984ApJ...283..694K}. It turns out that recent relativistic MHD simulations of the Crab Nebula show that such a magnetically dominated nebula is permitted\cite{2013MNRAS.431L..48P, 2013arXiv1310.2531P}, because kink-like instabilities diminish the magnetic hoop stress on the nebula, and hence preserve the morphology of the nebula even at high-$\sigma$ (Ref.~\onlinecite{1998ApJ...493..291B}). The dissipation of the magnetic energy should be done {\em in situ}, possibly via magnetic reconnection throughout the nebula provided that the kink instability (or other process) is efficient at creating current sheets. Reconnection may then be the primary particle acceleration process in the nebula since diffuse shock-acceleration is quenched for high-$\sigma$. The gamma-ray flares could be the smoking gun of the most spectacular episodes of magnetic dissipation in the nebula, analogous to saw-tooth crashes in tokamaks, and perhaps the first strong evidence of relativistic magnetic reconnection outside the Solar System.

\begin{acknowledgments}
BC acknowledges support from the Lyman Spitzer Jr. Fellowship awarded by the Department of Astrophysical Sciences at Princeton University, and the Max-Planck/Princeton Center for Plasma Physics. This work was also supported by NSF grant PHY-0903851, DOE Grants DE-SC0008409 and DE-SC0008655, NASA grant NNX12AP17G through the Fermi Guest Investigator Program. Numerical simulations were performed on the local CIPS computer cluster Verus, on Kraken at the National Institute for Computational Sciences (\url{www.nics.tennessee.edu/}) and on . This work also utilized the Janus supercomputer, which is supported by the National Science Foundation (award number CNS-0821794), the University of Colorado Boulder, the University of Colorado Denver, and the National Center for Atmospheric Research. The Janus supercomputer is operated by the University of Colorado Boulder.
\end{acknowledgments}

\bibliography{dpp2013}

\begin{thebibliography}{10}%
\makeatletter
\providecommand \@ifxundefined [1]{%
 \ifx #1\undefined \expandafter \@firstoftwo
 \else \expandafter \@secondoftwo
\fi
}%
\providecommand \@ifnum [1]{%
 \ifnum #1\expandafter \@firstoftwo
 \else \expandafter \@secondoftwo
\fi
}%
\providecommand \enquote [1]{``#1''}%
\providecommand \bibnamefont  [1]{#1}%
\providecommand \bibfnamefont [1]{#1}%
\providecommand \citenamefont [1]{#1}%
\providecommand\href[0]{\@sanitize\@href}%
\providecommand\@href[1]{\endgroup\@@startlink{#1}\endgroup\@@href}%
\providecommand\@@href[1]{#1\@@endlink}%
\providecommand \@sanitize [0]{\begingroup\catcode`\&12\catcode`\#12\relax}%
\@ifxundefined \pdfoutput {\@firstoftwo}{%
 \@ifnum{\z@=\pdfoutput}{\@firstoftwo}{\@secondoftwo}%
}{%
 \providecommand\@@startlink[1]{\leavevmode\special{html:<a href="#1">}}%
 \providecommand\@@endlink[0]{\special{html:</a>}}%
}{%
 \providecommand\@@startlink[1]{%
  \leavevmode
  \pdfstartlink
   attr{/Border[0 0 1 ]/H/I/C[0 1 1]}%
   user{/Subtype/Link/A<</Type/Action/S/URI/URI(#1)>>}%
  \relax
 }%
 \providecommand\@@endlink[0]{\pdfendlink}%
}%
\providecommand \url  [0]{\begingroup\@sanitize \@url }%
\providecommand \@url [1]{\endgroup\@href {#1}{\urlprefix}}%
\providecommand \urlprefix [0]{URL }%
\providecommand \Eprint[0]{\href }%
\@ifxundefined \urlstyle {%
  \providecommand \doi [1]{doi:\discretionary{}{}{}#1}%
}{%
  \providecommand \doi [0]{doi:\discretionary{}{}{}\begingroup
  \urlstyle{rm}\Url }%
}%
\providecommand \doibase [0]{http://dx.doi.org/}%
\providecommand \Doi[1]{\href{\doibase#1}}%
\providecommand \selectlanguage [0]{\@gobble}%
\providecommand \bibinfo [0]{\@secondoftwo}%
\providecommand \bibfield [0]{\@secondoftwo}%
\providecommand \translation [1]{[#1]}%
\providecommand \BibitemOpen[0]{}%
\providecommand \bibitemStop [0]{}%
\providecommand \bibitemNoStop [0]{.\EOS\space}%
\providecommand \EOS [0]{\spacefactor3000\relax}%
\providecommand \BibitemShut [1]{\csname bibitem#1\endcsname}%
\bibitem{1974MNRAS.167....1R}%
  \BibitemOpen
  \bibfield{author}{%
  \bibinfo {author} {\bibfnamefont{M.~J.}\ \bibnamefont{{Rees}}}\ and\ \bibinfo
  {author} {\bibfnamefont{J.~E.}\ \bibnamefont{{Gunn}}},\ }%
  \bibfield{title}{%
  \enquote{\bibinfo {title} {{The origin of the magnetic field and relativistic
  particles in the Crab Nebula}},}\ }%
  \bibfield{journal}{%
  \bibinfo {journal} {MNRAS}\ }%
  \textbf{\bibinfo {volume} {167}},\ \bibinfo {pages} {1--12} (\bibinfo {month}
  {Apr.}\ \bibinfo {year} {1974})\BibitemShut{NoStop}%
\bibitem{1984ApJ...283..694K}%
  \BibitemOpen
  \bibfield{author}{%
  \bibinfo {author} {\bibfnamefont{C.~F.}\ \bibnamefont{{Kennel}}}\ and\
  \bibinfo {author} {\bibfnamefont{F.~V.}\ \bibnamefont{{Coroniti}}},\ }%
  \bibfield{title}{%
  \enquote{\bibinfo {title} {{Confinement of the Crab pulsar's wind by its
  supernova remnant}},}\ }%
  \bibfield{journal}{%
  \Doi{10.1086/162356}{\bibinfo {journal} {ApJ}}\ }%
  \textbf{\bibinfo {volume} {283}},\ \bibinfo {pages} {694--709} (\bibinfo
  {month} {Aug.}\ \bibinfo {year} {1984})\BibitemShut{NoStop}%
\bibitem{2009ASSL..357..421K}%
  \BibitemOpen
  \bibfield{author}{%
  \bibinfo {author} {\bibfnamefont{J.~G.}\ \bibnamefont{{Kirk}}}, \bibinfo
  {author} {\bibfnamefont{Y.}~\bibnamefont{{Lyubarsky}}},\ and\ \bibinfo
  {author} {\bibfnamefont{J.}~\bibnamefont{{Petri}}},\ }%
  \enquote{\bibinfo {title} {{The Theory of Pulsar Winds and Nebulae}},}\ in\
  \Doi{10.1007/978-3-540-76965-1_16}{\emph{\bibinfo {booktitle} {Astrophysics
  and Space Science Library}}},\ \bibinfo {series} {Astrophysics and Space
  Science Library}, Vol.\ \bibinfo {volume} {357},\ \bibinfo {editor} {edited
  by\ \bibinfo {editor} {\bibfnamefont{W.}~\bibnamefont{{Becker}}}}\ (\bibinfo
  {year} {2009})\ p.\ \bibinfo {pages} {421},\
  \Eprint{http://arxiv.org/abs/astro-ph/0703116}{astro-ph/0703116}\BibitemShut%
{NoStop}%
\bibitem{2012SSRv..173..341A}%
  \BibitemOpen
  \bibfield{author}{%
  \bibinfo {author} {\bibfnamefont{J.}~\bibnamefont{{Arons}}},\ }%
  \bibfield{title}{%
  \enquote{\bibinfo {title} {{Pulsar Wind Nebulae as Cosmic Pevatrons: A
  Current Sheet's Tale}},}\ }%
  \bibfield{journal}{%
  \Doi{10.1007/s11214-012-9885-1}{\bibinfo {journal} {Space Sci. Rev.}}\ }%
  \textbf{\bibinfo {volume} {173}},\ \bibinfo {pages} {341--367} (\bibinfo
  {month} {Nov.}\ \bibinfo {year} {2012}),\
  \Eprint{http://arxiv.org/abs/1208.5787}{arXiv:1208.5787
  [astro-ph.HE]}\BibitemShut{NoStop}%
\bibitem{2013arXiv1309.7046B}%
  \BibitemOpen
  \bibfield{author}{%
  \bibinfo {author} {\bibfnamefont{R.}~\bibnamefont{{Buehler}}}\ and\ \bibinfo
  {author} {\bibfnamefont{R.}~\bibnamefont{{Blandford}}},\ }%
  \bibfield{title}{%
  \enquote{\bibinfo {title} {{The Crab pulsar wind nebula: our laboratory of
  the non-thermal Universe}},}\ }%
  \bibfield{journal}{%
  \bibinfo {journal} {ArXiv e-prints}}%
   (\bibinfo {month} {Sep.}\ \bibinfo {year} {2013}),\
  \Eprint{http://arxiv.org/abs/1309.7046}{arXiv:1309.7046
  [astro-ph.HE]}\BibitemShut{NoStop}%
\bibitem{1996MNRAS.278..525A}%
  \BibitemOpen
  \bibfield{author}{%
  \bibinfo {author} {\bibfnamefont{A.~M.}\ \bibnamefont{{Atoyan}}}\ and\
  \bibinfo {author} {\bibfnamefont{F.~A.}\ \bibnamefont{{Aharonian}}},\ }%
  \bibfield{title}{%
  \enquote{\bibinfo {title} {{On the mechanisms of gamma radiation in the Crab
  Nebula}},}\ }%
  \bibfield{journal}{%
  \bibinfo {journal} {MNRAS}\ }%
  \textbf{\bibinfo {volume} {278}},\ \bibinfo {pages} {525--541} (\bibinfo
  {month} {Jan.}\ \bibinfo {year} {1996})\BibitemShut{NoStop}%
\bibitem{2010A&A...523A...2M}%
  \BibitemOpen
  \bibfield{author}{%
  \bibinfo {author} {\bibfnamefont{M.}~\bibnamefont{{Meyer}}}, \bibinfo
  {author} {\bibfnamefont{D.}~\bibnamefont{{Horns}}},\ and\ \bibinfo {author}
  {\bibfnamefont{H.-S.}\ \bibnamefont{{Zechlin}}},\ }%
  \bibfield{title}{%
  \enquote{\bibinfo {title} {{The Crab Nebula as a standard candle in very
  high-energy astrophysics}},}\ }%
  \bibfield{journal}{%
  \Doi{10.1051/0004-6361/201014108}{\bibinfo {journal} {A\&A}}\ }%
  \textbf{\bibinfo {volume} {523}},\ \bibinfo {eid} {A2} (\bibinfo {month}
  {Nov.}\ \bibinfo {year} {2010}),\
  \Eprint{http://arxiv.org/abs/1008.4524}{arXiv:1008.4524
  [astro-ph.HE]}\BibitemShut{NoStop}%
\bibitem{2011Sci...331..736T}%
  \BibitemOpen
  \bibfield{author}{%
  \bibinfo {author} {\bibfnamefont{M.}~\bibnamefont{{Tavani}}}, \bibinfo
  {author} {\bibfnamefont{A.}~\bibnamefont{{Bulgarelli}}}, \bibinfo {author}
  {\bibfnamefont{V.}~\bibnamefont{{Vittorini}}}, \bibinfo {author}
  {\bibfnamefont{A.}~\bibnamefont{{Pellizzoni}}}, \bibinfo {author}
  {\bibfnamefont{E.}~\bibnamefont{{Striani}}}, \bibinfo {author}
  {\bibfnamefont{P.}~\bibnamefont{{Caraveo}}}, \bibinfo {author}
  {\bibfnamefont{M.~C.}\ \bibnamefont{{Weisskopf}}}, \bibinfo {author}
  {\bibfnamefont{A.}~\bibnamefont{{Tennant}}}, \bibinfo {author}
  {\bibfnamefont{G.}~\bibnamefont{{Pucella}}}, \bibinfo {author}
  {\bibfnamefont{A.}~\bibnamefont{{Trois}}}, \bibinfo {author}
  {\bibfnamefont{E.}~\bibnamefont{{Costa}}}, \bibinfo {author}
  {\bibfnamefont{Y.}~\bibnamefont{{Evangelista}}}, \bibinfo {author}
  {\bibfnamefont{C.}~\bibnamefont{{Pittori}}}, \bibinfo {author}
  {\bibfnamefont{F.}~\bibnamefont{{Verrecchia}}}, \bibinfo {author}
  {\bibfnamefont{E.}~\bibnamefont{{Del Monte}}}, \bibinfo {author}
  {\bibfnamefont{R.}~\bibnamefont{{Campana}}}, \bibinfo {author}
  {\bibfnamefont{M.}~\bibnamefont{{Pilia}}}, \bibinfo {author}
  {\bibfnamefont{A.}~\bibnamefont{{De Luca}}}, \bibinfo {author}
  {\bibfnamefont{I.}~\bibnamefont{{Donnarumma}}}, \bibinfo {author}
  {\bibfnamefont{D.}~\bibnamefont{{Horns}}}, \bibinfo {author}
  {\bibfnamefont{C.}~\bibnamefont{{Ferrigno}}}, \bibinfo {author}
  {\bibfnamefont{C.~O.}\ \bibnamefont{{Heinke}}}, \bibinfo {author}
  {\bibfnamefont{M.}~\bibnamefont{{Trifoglio}}}, \bibinfo {author}
  {\bibfnamefont{F.}~\bibnamefont{{Gianotti}}}, \bibinfo {author}
  {\bibfnamefont{S.}~\bibnamefont{{Vercellone}}}, \bibinfo {author}
  {\bibfnamefont{A.}~\bibnamefont{{Argan}}}, \bibinfo {author}
  {\bibfnamefont{G.}~\bibnamefont{{Barbiellini}}}, \bibinfo {author}
  {\bibfnamefont{P.~W.}\ \bibnamefont{{Cattaneo}}}, \bibinfo {author}
  {\bibfnamefont{A.~W.}\ \bibnamefont{{Chen}}}, \bibinfo {author}
  {\bibfnamefont{T.}~\bibnamefont{{Contessi}}}, \bibinfo {author}
  {\bibfnamefont{F.}~\bibnamefont{{D'Ammando}}}, \bibinfo {author}
  {\bibfnamefont{G.}~\bibnamefont{{DeParis}}}, \bibinfo {author}
  {\bibfnamefont{G.}~\bibnamefont{{Di Cocco}}}, \bibinfo {author}
  {\bibfnamefont{G.}~\bibnamefont{{Di Persio}}}, \bibinfo {author}
  {\bibfnamefont{M.}~\bibnamefont{{Feroci}}}, \bibinfo {author}
  {\bibfnamefont{A.}~\bibnamefont{{Ferrari}}}, \bibinfo {author}
  {\bibfnamefont{M.}~\bibnamefont{{Galli}}}, \bibinfo {author}
  {\bibfnamefont{A.}~\bibnamefont{{Giuliani}}}, \bibinfo {author}
  {\bibfnamefont{M.}~\bibnamefont{{Giusti}}}, \bibinfo {author}
  {\bibfnamefont{C.}~\bibnamefont{{Labanti}}}, \bibinfo {author}
  {\bibfnamefont{I.}~\bibnamefont{{Lapshov}}}, \bibinfo {author}
  {\bibfnamefont{F.}~\bibnamefont{{Lazzarotto}}}, \bibinfo {author}
  {\bibfnamefont{P.}~\bibnamefont{{Lipari}}}, \bibinfo {author}
  {\bibfnamefont{F.}~\bibnamefont{{Longo}}}, \bibinfo {author}
  {\bibfnamefont{F.}~\bibnamefont{{Fuschino}}}, \bibinfo {author}
  {\bibfnamefont{M.}~\bibnamefont{{Marisaldi}}}, \bibinfo {author}
  {\bibfnamefont{S.}~\bibnamefont{{Mereghetti}}}, \bibinfo {author}
  {\bibfnamefont{E.}~\bibnamefont{{Morelli}}}, \bibinfo {author}
  {\bibfnamefont{E.}~\bibnamefont{{Moretti}}}, \bibinfo {author}
  {\bibfnamefont{A.}~\bibnamefont{{Morselli}}}, \bibinfo {author}
  {\bibfnamefont{L.}~\bibnamefont{{Pacciani}}}, \bibinfo {author}
  {\bibfnamefont{F.}~\bibnamefont{{Perotti}}}, \bibinfo {author}
  {\bibfnamefont{G.}~\bibnamefont{{Piano}}}, \bibinfo {author}
  {\bibfnamefont{P.}~\bibnamefont{{Picozza}}}, \bibinfo {author}
  {\bibfnamefont{M.}~\bibnamefont{{Prest}}}, \bibinfo {author}
  {\bibfnamefont{M.}~\bibnamefont{{Rapisarda}}}, \bibinfo {author}
  {\bibfnamefont{A.}~\bibnamefont{{Rappoldi}}}, \bibinfo {author}
  {\bibfnamefont{A.}~\bibnamefont{{Rubini}}}, \bibinfo {author}
  {\bibfnamefont{S.}~\bibnamefont{{Sabatini}}}, \bibinfo {author}
  {\bibfnamefont{P.}~\bibnamefont{{Soffitta}}}, \bibinfo {author}
  {\bibfnamefont{E.}~\bibnamefont{{Vallazza}}}, \bibinfo {author}
  {\bibfnamefont{A.}~\bibnamefont{{Zambra}}}, \bibinfo {author}
  {\bibfnamefont{D.}~\bibnamefont{{Zanello}}}, \bibinfo {author}
  {\bibfnamefont{F.}~\bibnamefont{{Lucarelli}}}, \bibinfo {author}
  {\bibfnamefont{P.}~\bibnamefont{{Santolamazza}}}, \bibinfo {author}
  {\bibfnamefont{P.}~\bibnamefont{{Giommi}}}, \bibinfo {author}
  {\bibfnamefont{L.}~\bibnamefont{{Salotti}}},\ and\ \bibinfo {author}
  {\bibfnamefont{G.~F.}\ \bibnamefont{{Bignami}}},\ }%
  \bibfield{title}{%
  \enquote{\bibinfo {title} {{Discovery of Powerful Gamma-Ray Flares from the
  Crab Nebula}},}\ }%
  \bibfield{journal}{%
  \Doi{10.1126/science.1200083}{\bibinfo {journal} {Science}}\ }%
  \textbf{\bibinfo {volume} {331}},\ \bibinfo {pages} {736--} (\bibinfo {month}
  {Feb.}\ \bibinfo {year} {2011}),\
  \Eprint{http://arxiv.org/abs/1101.2311}{arXiv:1101.2311
  [astro-ph.HE]}\BibitemShut{NoStop}%
\bibitem{2011Sci...331..739A}%
  \BibitemOpen
  \bibfield{author}{%
  \bibinfo {author} {\bibfnamefont{A.~A.}\ \bibnamefont{{Abdo}}}, \bibinfo
  {author} {\bibfnamefont{M.}~\bibnamefont{{Ackermann}}}, \bibinfo {author}
  {\bibfnamefont{M.}~\bibnamefont{{Ajello}}}, \bibinfo {author}
  {\bibfnamefont{A.}~\bibnamefont{{Allafort}}}, \bibinfo {author}
  {\bibfnamefont{L.}~\bibnamefont{{Baldini}}}, \bibinfo {author}
  {\bibfnamefont{J.}~\bibnamefont{{Ballet}}}, \bibinfo {author}
  {\bibfnamefont{G.}~\bibnamefont{{Barbiellini}}}, \bibinfo {author}
  {\bibfnamefont{D.}~\bibnamefont{{Bastieri}}}, \bibinfo {author}
  {\bibfnamefont{K.}~\bibnamefont{{Bechtol}}}, \bibinfo {author}
  {\bibfnamefont{R.}~\bibnamefont{{Bellazzini}}}, \bibinfo {author}
  {\bibfnamefont{B.}~\bibnamefont{{Berenji}}}, \bibinfo {author}
  {\bibfnamefont{R.~D.}\ \bibnamefont{{Blandford}}}, \bibinfo {author}
  {\bibfnamefont{E.~D.}\ \bibnamefont{{Bloom}}}, \bibinfo {author}
  {\bibfnamefont{E.}~\bibnamefont{{Bonamente}}}, \bibinfo {author}
  {\bibfnamefont{A.~W.}\ \bibnamefont{{Borgland}}}, \bibinfo {author}
  {\bibfnamefont{A.}~\bibnamefont{{Bouvier}}}, \bibinfo {author}
  {\bibfnamefont{T.~J.}\ \bibnamefont{{Brandt}}}, \bibinfo {author}
  {\bibfnamefont{J.}~\bibnamefont{{Bregeon}}}, \bibinfo {author}
  {\bibfnamefont{A.}~\bibnamefont{{Brez}}}, \bibinfo {author}
  {\bibfnamefont{M.}~\bibnamefont{{Brigida}}}, \bibinfo {author}
  {\bibfnamefont{P.}~\bibnamefont{{Bruel}}}, \bibinfo {author}
  {\bibfnamefont{R.}~\bibnamefont{{Buehler}}}, \bibinfo {author}
  {\bibfnamefont{S.}~\bibnamefont{{Buson}}}, \bibinfo {author}
  {\bibfnamefont{G.~A.}\ \bibnamefont{{Caliandro}}}, \bibinfo {author}
  {\bibfnamefont{R.~A.}\ \bibnamefont{{Cameron}}}, \bibinfo {author}
  {\bibfnamefont{A.}~\bibnamefont{{Cannon}}}, \bibinfo {author}
  {\bibfnamefont{P.~A.}\ \bibnamefont{{Caraveo}}}, \bibinfo {author}
  {\bibfnamefont{J.~M.}\ \bibnamefont{{Casandjian}}}, \bibinfo {author}
  {\bibfnamefont{{\"O}.}~\bibnamefont{{{\c C}elik}}}, \bibinfo {author}
  {\bibfnamefont{E.}~\bibnamefont{{Charles}}}, \bibinfo {author}
  {\bibfnamefont{A.}~\bibnamefont{{Chekhtman}}}, \bibinfo {author}
  {\bibfnamefont{C.~C.}\ \bibnamefont{{Cheung}}}, \bibinfo {author}
  {\bibfnamefont{J.}~\bibnamefont{{Chiang}}}, \bibinfo {author}
  {\bibfnamefont{S.}~\bibnamefont{{Ciprini}}}, \bibinfo {author}
  {\bibfnamefont{R.}~\bibnamefont{{Claus}}}, \bibinfo {author}
  {\bibfnamefont{J.}~\bibnamefont{{Cohen-Tanugi}}}, \bibinfo {author}
  {\bibfnamefont{L.}~\bibnamefont{{Costamante}}}, \bibinfo {author}
  {\bibfnamefont{S.}~\bibnamefont{{Cutini}}}, \bibinfo {author}
  {\bibfnamefont{F.}~\bibnamefont{{D'Ammando}}}, \bibinfo {author}
  {\bibfnamefont{C.~D.}\ \bibnamefont{{Dermer}}}, \bibinfo {author}
  {\bibfnamefont{A.}~\bibnamefont{{de Angelis}}}, \bibinfo {author}
  {\bibfnamefont{A.}~\bibnamefont{{de Luca}}}, \bibinfo {author}
  {\bibfnamefont{F.}~\bibnamefont{{de Palma}}}, \bibinfo {author}
  {\bibfnamefont{S.~W.}\ \bibnamefont{{Digel}}}, \bibinfo {author}
  {\bibfnamefont{E.}~\bibnamefont{{do Couto e Silva}}}, \bibinfo {author}
  {\bibfnamefont{P.~S.}\ \bibnamefont{{Drell}}}, \bibinfo {author}
  {\bibfnamefont{A.}~\bibnamefont{{Drlica-Wagner}}}, \bibinfo {author}
  {\bibfnamefont{R.}~\bibnamefont{{Dubois}}}, \bibinfo {author}
  {\bibfnamefont{D.}~\bibnamefont{{Dumora}}}, \bibinfo {author}
  {\bibfnamefont{C.}~\bibnamefont{{Favuzzi}}}, \bibinfo {author}
  {\bibfnamefont{S.~J.}\ \bibnamefont{{Fegan}}}, \bibinfo {author}
  {\bibfnamefont{E.~C.}\ \bibnamefont{{Ferrara}}}, \bibinfo {author}
  {\bibfnamefont{W.~B.}\ \bibnamefont{{Focke}}}, \bibinfo {author}
  {\bibfnamefont{P.}~\bibnamefont{{Fortin}}}, \bibinfo {author}
  {\bibfnamefont{M.}~\bibnamefont{{Frailis}}}, \bibinfo {author}
  {\bibfnamefont{Y.}~\bibnamefont{{Fukazawa}}}, \bibinfo {author}
  {\bibfnamefont{S.}~\bibnamefont{{Funk}}}, \bibinfo {author}
  {\bibfnamefont{P.}~\bibnamefont{{Fusco}}}, \bibinfo {author}
  {\bibfnamefont{F.}~\bibnamefont{{Gargano}}}, \bibinfo {author}
  {\bibfnamefont{D.}~\bibnamefont{{Gasparrini}}}, \bibinfo {author}
  {\bibfnamefont{N.}~\bibnamefont{{Gehrels}}}, \bibinfo {author}
  {\bibfnamefont{S.}~\bibnamefont{{Germani}}}, \bibinfo {author}
  {\bibfnamefont{N.}~\bibnamefont{{Giglietto}}}, \bibinfo {author}
  {\bibfnamefont{F.}~\bibnamefont{{Giordano}}}, \bibinfo {author}
  {\bibfnamefont{M.}~\bibnamefont{{Giroletti}}}, \bibinfo {author}
  {\bibfnamefont{T.}~\bibnamefont{{Glanzman}}}, \bibinfo {author}
  {\bibfnamefont{G.}~\bibnamefont{{Godfrey}}}, \bibinfo {author}
  {\bibfnamefont{I.~A.}\ \bibnamefont{{Grenier}}}, \bibinfo {author}
  {\bibfnamefont{M.-H.}\ \bibnamefont{{Grondin}}}, \bibinfo {author}
  {\bibfnamefont{J.~E.}\ \bibnamefont{{Grove}}}, \bibinfo {author}
  {\bibfnamefont{S.}~\bibnamefont{{Guiriec}}}, \bibinfo {author}
  {\bibfnamefont{D.}~\bibnamefont{{Hadasch}}}, \bibinfo {author}
  {\bibfnamefont{Y.}~\bibnamefont{{Hanabata}}}, \bibinfo {author}
  {\bibfnamefont{A.~K.}\ \bibnamefont{{Harding}}}, \bibinfo {author}
  {\bibfnamefont{K.}~\bibnamefont{{Hayashi}}}, \bibinfo {author}
  {\bibfnamefont{M.}~\bibnamefont{{Hayashida}}}, \bibinfo {author}
  {\bibfnamefont{E.}~\bibnamefont{{Hays}}}, \bibinfo {author}
  {\bibfnamefont{D.}~\bibnamefont{{Horan}}}, \bibinfo {author}
  {\bibfnamefont{R.}~\bibnamefont{{Itoh}}}, \bibinfo {author}
  {\bibfnamefont{G.}~\bibnamefont{{J{\'o}hannesson}}}, \bibinfo {author}
  {\bibfnamefont{A.~S.}\ \bibnamefont{{Johnson}}}, \bibinfo {author}
  {\bibfnamefont{T.~J.}\ \bibnamefont{{Johnson}}}, \bibinfo {author}
  {\bibfnamefont{D.}~\bibnamefont{{Khangulyan}}}, \bibinfo {author}
  {\bibfnamefont{T.}~\bibnamefont{{Kamae}}}, \bibinfo {author}
  {\bibfnamefont{H.}~\bibnamefont{{Katagiri}}}, \bibinfo {author}
  {\bibfnamefont{J.}~\bibnamefont{{Kataoka}}}, \bibinfo {author}
  {\bibfnamefont{M.}~\bibnamefont{{Kerr}}}, \bibinfo {author}
  {\bibfnamefont{J.}~\bibnamefont{{Kn{\"o}dlseder}}}, \bibinfo {author}
  {\bibfnamefont{M.}~\bibnamefont{{Kuss}}}, \bibinfo {author}
  {\bibfnamefont{J.}~\bibnamefont{{Lande}}}, \bibinfo {author}
  {\bibfnamefont{L.}~\bibnamefont{{Latronico}}}, \bibinfo {author}
  {\bibfnamefont{S.-H.}\ \bibnamefont{{Lee}}}, \bibinfo {author}
  {\bibfnamefont{M.}~\bibnamefont{{Lemoine-Goumard}}}, \bibinfo {author}
  {\bibfnamefont{F.}~\bibnamefont{{Longo}}}, \bibinfo {author}
  {\bibfnamefont{F.}~\bibnamefont{{Loparco}}}, \bibinfo {author}
  {\bibfnamefont{P.}~\bibnamefont{{Lubrano}}}, \bibinfo {author}
  {\bibfnamefont{G.~M.}\ \bibnamefont{{Madejski}}}, \bibinfo {author}
  {\bibfnamefont{A.}~\bibnamefont{{Makeev}}}, \bibinfo {author}
  {\bibfnamefont{M.}~\bibnamefont{{Marelli}}}, \bibinfo {author}
  {\bibfnamefont{M.~N.}\ \bibnamefont{{Mazziotta}}}, \bibinfo {author}
  {\bibfnamefont{J.~E.}\ \bibnamefont{{McEnery}}}, \bibinfo {author}
  {\bibfnamefont{P.~F.}\ \bibnamefont{{Michelson}}}, \bibinfo {author}
  {\bibfnamefont{W.}~\bibnamefont{{Mitthumsiri}}}, \bibinfo {author}
  {\bibfnamefont{T.}~\bibnamefont{{Mizuno}}}, \bibinfo {author}
  {\bibfnamefont{A.~A.}\ \bibnamefont{{Moiseev}}}, \bibinfo {author}
  {\bibfnamefont{C.}~\bibnamefont{{Monte}}}, \bibinfo {author}
  {\bibfnamefont{M.~E.}\ \bibnamefont{{Monzani}}}, \bibinfo {author}
  {\bibfnamefont{A.}~\bibnamefont{{Morselli}}}, \bibinfo {author}
  {\bibfnamefont{I.~V.}\ \bibnamefont{{Moskalenko}}}, \bibinfo {author}
  {\bibfnamefont{S.}~\bibnamefont{{Murgia}}}, \bibinfo {author}
  {\bibfnamefont{T.}~\bibnamefont{{Nakamori}}}, \bibinfo {author}
  {\bibfnamefont{M.}~\bibnamefont{{Naumann-Godo}}}, \bibinfo {author}
  {\bibfnamefont{P.~L.}\ \bibnamefont{{Nolan}}}, \bibinfo {author}
  {\bibfnamefont{J.~P.}\ \bibnamefont{{Norris}}}, \bibinfo {author}
  {\bibfnamefont{E.}~\bibnamefont{{Nuss}}}, \bibinfo {author}
  {\bibfnamefont{T.}~\bibnamefont{{Ohsugi}}}, \bibinfo {author}
  {\bibfnamefont{A.}~\bibnamefont{{Okumura}}}, \bibinfo {author}
  {\bibfnamefont{N.}~\bibnamefont{{Omodei}}}, \bibinfo {author}
  {\bibfnamefont{J.~F.}\ \bibnamefont{{Ormes}}}, \bibinfo {author}
  {\bibfnamefont{M.}~\bibnamefont{{Ozaki}}}, \bibinfo {author}
  {\bibfnamefont{D.}~\bibnamefont{{Paneque}}}, \bibinfo {author}
  {\bibfnamefont{D.}~\bibnamefont{{Parent}}}, \bibinfo {author}
  {\bibfnamefont{V.}~\bibnamefont{{Pelassa}}}, \bibinfo {author}
  {\bibfnamefont{M.}~\bibnamefont{{Pepe}}}, \bibinfo {author}
  {\bibfnamefont{M.}~\bibnamefont{{Pesce-Rollins}}}, \bibinfo {author}
  {\bibfnamefont{M.}~\bibnamefont{{Pierbattista}}}, \bibinfo {author}
  {\bibfnamefont{F.}~\bibnamefont{{Piron}}}, \bibinfo {author}
  {\bibfnamefont{T.~A.}\ \bibnamefont{{Porter}}}, \bibinfo {author}
  {\bibfnamefont{S.}~\bibnamefont{{Rain{\`o}}}}, \bibinfo {author}
  {\bibfnamefont{R.}~\bibnamefont{{Rando}}}, \bibinfo {author}
  {\bibfnamefont{P.~S.}\ \bibnamefont{{Ray}}}, \bibinfo {author}
  {\bibfnamefont{M.}~\bibnamefont{{Razzano}}}, \bibinfo {author}
  {\bibfnamefont{A.}~\bibnamefont{{Reimer}}}, \bibinfo {author}
  {\bibfnamefont{O.}~\bibnamefont{{Reimer}}}, \bibinfo {author}
  {\bibfnamefont{T.}~\bibnamefont{{Reposeur}}}, \bibinfo {author}
  {\bibfnamefont{S.}~\bibnamefont{{Ritz}}}, \bibinfo {author}
  {\bibfnamefont{R.~W.}\ \bibnamefont{{Romani}}}, \bibinfo {author}
  {\bibfnamefont{H.~F.-W.}\ \bibnamefont{{Sadrozinski}}}, \bibinfo {author}
  {\bibfnamefont{D.}~\bibnamefont{{Sanchez}}}, \bibinfo {author}
  {\bibfnamefont{P.~M.~S.}\ \bibnamefont{{Parkinson}}}, \bibinfo {author}
  {\bibfnamefont{J.~D.}\ \bibnamefont{{Scargle}}}, \bibinfo {author}
  {\bibfnamefont{T.~L.}\ \bibnamefont{{Schalk}}}, \bibinfo {author}
  {\bibfnamefont{C.}~\bibnamefont{{Sgr{\`o}}}}, \bibinfo {author}
  {\bibfnamefont{E.~J.}\ \bibnamefont{{Siskind}}}, \bibinfo {author}
  {\bibfnamefont{P.~D.}\ \bibnamefont{{Smith}}}, \bibinfo {author}
  {\bibfnamefont{G.}~\bibnamefont{{Spandre}}}, \bibinfo {author}
  {\bibfnamefont{P.}~\bibnamefont{{Spinelli}}}, \bibinfo {author}
  {\bibfnamefont{M.~S.}\ \bibnamefont{{Strickman}}}, \bibinfo {author}
  {\bibfnamefont{D.~J.}\ \bibnamefont{{Suson}}}, \bibinfo {author}
  {\bibfnamefont{H.}~\bibnamefont{{Takahashi}}}, \bibinfo {author}
  {\bibfnamefont{T.}~\bibnamefont{{Takahashi}}}, \bibinfo {author}
  {\bibfnamefont{T.}~\bibnamefont{{Tanaka}}}, \bibinfo {author}
  {\bibfnamefont{J.~B.}\ \bibnamefont{{Thayer}}}, \bibinfo {author}
  {\bibfnamefont{D.~J.}\ \bibnamefont{{Thompson}}}, \bibinfo {author}
  {\bibfnamefont{L.}~\bibnamefont{{Tibaldo}}}, \bibinfo {author}
  {\bibfnamefont{D.~F.}\ \bibnamefont{{Torres}}}, \bibinfo {author}
  {\bibfnamefont{G.}~\bibnamefont{{Tosti}}}, \bibinfo {author}
  {\bibfnamefont{A.}~\bibnamefont{{Tramacere}}}, \bibinfo {author}
  {\bibfnamefont{E.}~\bibnamefont{{Troja}}}, \bibinfo {author}
  {\bibfnamefont{Y.}~\bibnamefont{{Uchiyama}}}, \bibinfo {author}
  {\bibfnamefont{J.}~\bibnamefont{{Vandenbroucke}}}, \bibinfo {author}
  {\bibfnamefont{V.}~\bibnamefont{{Vasileiou}}}, \bibinfo {author}
  {\bibfnamefont{G.}~\bibnamefont{{Vianello}}}, \bibinfo {author}
  {\bibfnamefont{V.}~\bibnamefont{{Vitale}}}, \bibinfo {author}
  {\bibfnamefont{P.}~\bibnamefont{{Wang}}}, \bibinfo {author}
  {\bibfnamefont{K.~S.}\ \bibnamefont{{Wood}}}, \bibinfo {author}
  {\bibfnamefont{Z.}~\bibnamefont{{Yang}}},\ and\ \bibinfo {author}
  {\bibfnamefont{M.}~\bibnamefont{{Ziegler}}},\ }%
  \bibfield{title}{%
  \enquote{\bibinfo {title} {{Gamma-Ray Flares from the Crab Nebula}},}\ }%
  \bibfield{journal}{%
  \Doi{10.1126/science.1199705}{\bibinfo {journal} {Science}}\ }%
  \textbf{\bibinfo {volume} {331}},\ \bibinfo {pages} {739--} (\bibinfo {month}
  {Feb.}\ \bibinfo {year} {2011}),\
  \Eprint{http://arxiv.org/abs/1011.3855}{arXiv:1011.3855
  [astro-ph.HE]}\BibitemShut{NoStop}%
\bibitem{2011A&A...527L...4B}%
  \BibitemOpen
  \bibfield{author}{%
  \bibinfo {author} {\bibfnamefont{M.}~\bibnamefont{{Balbo}}}, \bibinfo
  {author} {\bibfnamefont{R.}~\bibnamefont{{Walter}}}, \bibinfo {author}
  {\bibfnamefont{C.}~\bibnamefont{{Ferrigno}}},\ and\ \bibinfo {author}
  {\bibfnamefont{P.}~\bibnamefont{{Bordas}}},\ }%
  \bibfield{title}{%
  \enquote{\bibinfo {title} {{Twelve-hour spikes from the Crab Pevatron}},}\ }%
  \bibfield{journal}{%
  \Doi{10.1051/0004-6361/201015980}{\bibinfo {journal} {A\&A}}\ }%
  \textbf{\bibinfo {volume} {527}},\ \bibinfo {eid} {L4} (\bibinfo {month}
  {Mar.}\ \bibinfo {year} {2011}),\
  \Eprint{http://arxiv.org/abs/1012.3397}{arXiv:1012.3397
  [astro-ph.HE]}\BibitemShut{NoStop}%
\bibitem{2011ApJ...741L...5S}%
  \BibitemOpen
  \bibfield{author}{%
  \bibinfo {author} {\bibfnamefont{E.}~\bibnamefont{{Striani}}}, \bibinfo
  {author} {\bibfnamefont{M.}~\bibnamefont{{Tavani}}}, \bibinfo {author}
  {\bibfnamefont{G.}~\bibnamefont{{Piano}}}, \bibinfo {author}
  {\bibfnamefont{I.}~\bibnamefont{{Donnarumma}}}, \bibinfo {author}
  {\bibfnamefont{G.}~\bibnamefont{{Pucella}}}, \bibinfo {author}
  {\bibfnamefont{V.}~\bibnamefont{{Vittorini}}}, \bibinfo {author}
  {\bibfnamefont{A.}~\bibnamefont{{Bulgarelli}}}, \bibinfo {author}
  {\bibfnamefont{A.}~\bibnamefont{{Trois}}}, \bibinfo {author}
  {\bibfnamefont{C.}~\bibnamefont{{Pittori}}}, \bibinfo {author}
  {\bibfnamefont{F.}~\bibnamefont{{Verrecchia}}}, \bibinfo {author}
  {\bibfnamefont{E.}~\bibnamefont{{Costa}}}, \bibinfo {author}
  {\bibfnamefont{M.}~\bibnamefont{{Weisskopf}}}, \bibinfo {author}
  {\bibfnamefont{A.}~\bibnamefont{{Tennant}}}, \bibinfo {author}
  {\bibfnamefont{A.}~\bibnamefont{{Argan}}}, \bibinfo {author}
  {\bibfnamefont{G.}~\bibnamefont{{Barbiellini}}}, \bibinfo {author}
  {\bibfnamefont{P.}~\bibnamefont{{Caraveo}}}, \bibinfo {author}
  {\bibfnamefont{M.}~\bibnamefont{{Cardillo}}}, \bibinfo {author}
  {\bibfnamefont{P.~W.}\ \bibnamefont{{Cattaneo}}}, \bibinfo {author}
  {\bibfnamefont{A.~W.}\ \bibnamefont{{Chen}}}, \bibinfo {author}
  {\bibfnamefont{G.}~\bibnamefont{{De Paris}}}, \bibinfo {author}
  {\bibfnamefont{E.}~\bibnamefont{{Del Monte}}}, \bibinfo {author}
  {\bibfnamefont{G.}~\bibnamefont{{Di Cocco}}}, \bibinfo {author}
  {\bibfnamefont{Y.}~\bibnamefont{{Evangelista}}}, \bibinfo {author}
  {\bibfnamefont{A.}~\bibnamefont{{Ferrari}}}, \bibinfo {author}
  {\bibfnamefont{M.}~\bibnamefont{{Feroci}}}, \bibinfo {author}
  {\bibfnamefont{F.}~\bibnamefont{{Fuschino}}}, \bibinfo {author}
  {\bibfnamefont{M.}~\bibnamefont{{Galli}}}, \bibinfo {author}
  {\bibfnamefont{F.}~\bibnamefont{{Gianotti}}}, \bibinfo {author}
  {\bibfnamefont{A.}~\bibnamefont{{Giuliani}}}, \bibinfo {author}
  {\bibfnamefont{C.}~\bibnamefont{{Labanti}}}, \bibinfo {author}
  {\bibfnamefont{I.}~\bibnamefont{{Lapshov}}}, \bibinfo {author}
  {\bibfnamefont{F.}~\bibnamefont{{Lazzarotto}}}, \bibinfo {author}
  {\bibfnamefont{F.}~\bibnamefont{{Longo}}}, \bibinfo {author}
  {\bibfnamefont{M.}~\bibnamefont{{Marisaldi}}}, \bibinfo {author}
  {\bibfnamefont{S.}~\bibnamefont{{Mereghetti}}}, \bibinfo {author}
  {\bibfnamefont{A.}~\bibnamefont{{Morselli}}}, \bibinfo {author}
  {\bibfnamefont{L.}~\bibnamefont{{Pacciani}}}, \bibinfo {author}
  {\bibfnamefont{A.}~\bibnamefont{{Pellizzoni}}}, \bibinfo {author}
  {\bibfnamefont{F.}~\bibnamefont{{Perotti}}}, \bibinfo {author}
  {\bibfnamefont{P.}~\bibnamefont{{Picozza}}}, \bibinfo {author}
  {\bibfnamefont{M.}~\bibnamefont{{Pilia}}}, \bibinfo {author}
  {\bibfnamefont{M.}~\bibnamefont{{Rapisarda}}}, \bibinfo {author}
  {\bibfnamefont{A.}~\bibnamefont{{Rappoldi}}}, \bibinfo {author}
  {\bibfnamefont{S.}~\bibnamefont{{Sabatini}}}, \bibinfo {author}
  {\bibfnamefont{P.}~\bibnamefont{{Soffitta}}}, \bibinfo {author}
  {\bibfnamefont{M.}~\bibnamefont{{Trifoglio}}}, \bibinfo {author}
  {\bibfnamefont{S.}~\bibnamefont{{Vercellone}}}, \bibinfo {author}
  {\bibfnamefont{F.}~\bibnamefont{{Lucarelli}}}, \bibinfo {author}
  {\bibfnamefont{P.}~\bibnamefont{{Santolamazza}}},\ and\ \bibinfo {author}
  {\bibfnamefont{P.}~\bibnamefont{{Giommi}}},\ }%
  \bibfield{title}{%
  \enquote{\bibinfo {title} {{The Crab Nebula Super-flare in 2011 April:
  Extremely Fast Particle Acceleration and Gamma-Ray Emission}},}\ }%
  \bibfield{journal}{%
  \Doi{10.1088/2041-8205/741/1/L5}{\bibinfo {journal} {ApJL}}\ }%
  \textbf{\bibinfo {volume} {741}},\ \bibinfo {eid} {L5} (\bibinfo {month}
  {Nov.}\ \bibinfo {year} {2011}),\
  \Eprint{http://arxiv.org/abs/1105.5028}{arXiv:1105.5028
  [astro-ph.HE]}\BibitemShut{NoStop}%
\bibitem{2012ApJ...749...26B}%
  \BibitemOpen
  \bibfield{author}{%
  \bibinfo {author} {\bibfnamefont{R.}~\bibnamefont{{Buehler}}}, \bibinfo
  {author} {\bibfnamefont{J.~D.}\ \bibnamefont{{Scargle}}}, \bibinfo {author}
  {\bibfnamefont{R.~D.}\ \bibnamefont{{Blandford}}}, \bibinfo {author}
  {\bibfnamefont{L.}~\bibnamefont{{Baldini}}}, \bibinfo {author}
  {\bibfnamefont{M.~G.}\ \bibnamefont{{Baring}}}, \bibinfo {author}
  {\bibfnamefont{A.}~\bibnamefont{{Belfiore}}}, \bibinfo {author}
  {\bibfnamefont{E.}~\bibnamefont{{Charles}}}, \bibinfo {author}
  {\bibfnamefont{J.}~\bibnamefont{{Chiang}}}, \bibinfo {author}
  {\bibfnamefont{F.}~\bibnamefont{{D'Ammando}}}, \bibinfo {author}
  {\bibfnamefont{C.~D.}\ \bibnamefont{{Dermer}}}, \bibinfo {author}
  {\bibfnamefont{S.}~\bibnamefont{{Funk}}}, \bibinfo {author}
  {\bibfnamefont{J.~E.}\ \bibnamefont{{Grove}}}, \bibinfo {author}
  {\bibfnamefont{A.~K.}\ \bibnamefont{{Harding}}}, \bibinfo {author}
  {\bibfnamefont{E.}~\bibnamefont{{Hays}}}, \bibinfo {author}
  {\bibfnamefont{M.}~\bibnamefont{{Kerr}}}, \bibinfo {author}
  {\bibfnamefont{F.}~\bibnamefont{{Massaro}}}, \bibinfo {author}
  {\bibfnamefont{M.~N.}\ \bibnamefont{{Mazziotta}}}, \bibinfo {author}
  {\bibfnamefont{R.~W.}\ \bibnamefont{{Romani}}}, \bibinfo {author}
  {\bibfnamefont{P.~M.}\ \bibnamefont{{Saz Parkinson}}}, \bibinfo {author}
  {\bibfnamefont{A.~F.}\ \bibnamefont{{Tennant}}},\ and\ \bibinfo {author}
  {\bibfnamefont{M.~C.}\ \bibnamefont{{Weisskopf}}},\ }%
  \bibfield{title}{%
  \enquote{\bibinfo {title} {{Gamma-Ray Activity in the Crab Nebula: The
  Exceptional Flare of 2011 April}},}\ }%
  \bibfield{journal}{%
  \Doi{10.1088/0004-637X/749/1/26}{\bibinfo {journal} {ApJ}}\ }%
  \textbf{\bibinfo {volume} {749}},\ \bibinfo {eid} {26} (\bibinfo {month}
  {Apr.}\ \bibinfo {year} {2012}),\
  \Eprint{http://arxiv.org/abs/1112.1979}{arXiv:1112.1979
  [astro-ph.HE]}\BibitemShut{NoStop}%
\bibitem{2013ApJ...765...52S}%
  \BibitemOpen
  \bibfield{author}{%
  \bibinfo {author} {\bibfnamefont{E.}~\bibnamefont{{Striani}}}, \bibinfo
  {author} {\bibfnamefont{M.}~\bibnamefont{{Tavani}}}, \bibinfo {author}
  {\bibfnamefont{V.}~\bibnamefont{{Vittorini}}}, \bibinfo {author}
  {\bibfnamefont{I.}~\bibnamefont{{Donnarumma}}}, \bibinfo {author}
  {\bibfnamefont{A.}~\bibnamefont{{Giuliani}}}, \bibinfo {author}
  {\bibfnamefont{G.}~\bibnamefont{{Pucella}}}, \bibinfo {author}
  {\bibfnamefont{A.}~\bibnamefont{{Argan}}}, \bibinfo {author}
  {\bibfnamefont{A.}~\bibnamefont{{Bulgarelli}}}, \bibinfo {author}
  {\bibfnamefont{S.}~\bibnamefont{{Colafrancesco}}}, \bibinfo {author}
  {\bibfnamefont{M.}~\bibnamefont{{Cardillo}}}, \bibinfo {author}
  {\bibfnamefont{E.}~\bibnamefont{{Costa}}}, \bibinfo {author}
  {\bibfnamefont{E.}~\bibnamefont{{Del Monte}}}, \bibinfo {author}
  {\bibfnamefont{A.}~\bibnamefont{{Ferrari}}}, \bibinfo {author}
  {\bibfnamefont{S.}~\bibnamefont{{Mereghetti}}}, \bibinfo {author}
  {\bibfnamefont{L.}~\bibnamefont{{Pacciani}}}, \bibinfo {author}
  {\bibfnamefont{A.}~\bibnamefont{{Pellizzoni}}}, \bibinfo {author}
  {\bibfnamefont{G.}~\bibnamefont{{Piano}}}, \bibinfo {author}
  {\bibfnamefont{C.}~\bibnamefont{{Pittori}}}, \bibinfo {author}
  {\bibfnamefont{M.}~\bibnamefont{{Rapisarda}}}, \bibinfo {author}
  {\bibfnamefont{S.}~\bibnamefont{{Sabatini}}}, \bibinfo {author}
  {\bibfnamefont{P.}~\bibnamefont{{Soffitta}}}, \bibinfo {author}
  {\bibfnamefont{M.}~\bibnamefont{{Trifoglio}}}, \bibinfo {author}
  {\bibfnamefont{A.}~\bibnamefont{{Trois}}}, \bibinfo {author}
  {\bibfnamefont{S.}~\bibnamefont{{Vercellone}}},\ and\ \bibinfo {author}
  {\bibfnamefont{F.}~\bibnamefont{{Verrecchia}}},\ }%
  \bibfield{title}{%
  \enquote{\bibinfo {title} {{Variable Gamma-Ray Emission from the Crab Nebula:
  Short Flares and Long ''Waves''}},}\ }%
  \bibfield{journal}{%
  \Doi{10.1088/0004-637X/765/1/52}{\bibinfo {journal} {ApJ}}\ }%
  \textbf{\bibinfo {volume} {765}},\ \bibinfo {eid} {52} (\bibinfo {month}
  {Mar.}\ \bibinfo {year} {2013}),\
  \Eprint{http://arxiv.org/abs/1302.4342}{arXiv:1302.4342
  [astro-ph.HE]}\BibitemShut{NoStop}%
\bibitem{2013ApJ...775L..37M}%
  \BibitemOpen
  \bibfield{author}{%
  \bibinfo {author} {\bibfnamefont{M.}~\bibnamefont{{Mayer}}}, \bibinfo
  {author} {\bibfnamefont{R.}~\bibnamefont{{Buehler}}}, \bibinfo {author}
  {\bibfnamefont{E.}~\bibnamefont{{Hays}}}, \bibinfo {author}
  {\bibfnamefont{C.~C.}\ \bibnamefont{{Cheung}}}, \bibinfo {author}
  {\bibfnamefont{M.~S.}\ \bibnamefont{{Dutka}}}, \bibinfo {author}
  {\bibfnamefont{J.~E.}\ \bibnamefont{{Grove}}}, \bibinfo {author}
  {\bibfnamefont{M.}~\bibnamefont{{Kerr}}},\ and\ \bibinfo {author}
  {\bibfnamefont{R.}~\bibnamefont{{Ojha}}},\ }%
  \bibfield{title}{%
  \enquote{\bibinfo {title} {{Rapid Gamma-Ray Flux Variability during the 2013
  March Crab Nebula Flare}},}\ }%
  \bibfield{journal}{%
  \Doi{10.1088/2041-8205/775/2/L37}{\bibinfo {journal} {ApJL}}\ }%
  \textbf{\bibinfo {volume} {775}},\ \bibinfo {eid} {L37} (\bibinfo {month}
  {Oct.}\ \bibinfo {year} {2013}),\
  \Eprint{http://arxiv.org/abs/1308.6698}{arXiv:1308.6698
  [astro-ph.HE]}\BibitemShut{NoStop}%
\bibitem{2013ATel.5485....1B}%
  \BibitemOpen
  \bibfield{author}{%
  \bibinfo {author} {\bibfnamefont{S.}~\bibnamefont{{Buson}}}, \bibinfo
  {author} {\bibfnamefont{R.}~\bibnamefont{{Buehler}}},\ and\ \bibinfo {author}
  {\bibfnamefont{E.}~\bibnamefont{{Hays}}},\ }%
  \bibfield{title}{%
  \enquote{\bibinfo {title} {{Fermi LAT detection of enhanced gamma-ray
  emission from the Crab Nebula region}},}\ }%
  \bibfield{journal}{%
  \bibinfo {journal} {The Astronomer's Telegram}\ }%
  \textbf{\bibinfo {volume} {5485}},\ \bibinfo {pages} {1} (\bibinfo {month}
  {Oct.}\ \bibinfo {year} {2013})\BibitemShut{NoStop}%
\bibitem{2013ApJ...763..131T}%
  \BibitemOpen
  \bibfield{author}{%
  \bibinfo {author} {\bibfnamefont{Y.}~\bibnamefont{{Teraki}}}\ and\ \bibinfo
  {author} {\bibfnamefont{F.}~\bibnamefont{{Takahara}}},\ }%
  \bibfield{title}{%
  \enquote{\bibinfo {title} {{Jitter Radiation Model of the Crab Gamma-Ray
  Flares}},}\ }%
  \bibfield{journal}{%
  \Doi{10.1088/0004-637X/763/2/131}{\bibinfo {journal} {ApJ}}\ }%
  \textbf{\bibinfo {volume} {763}},\ \bibinfo {eid} {131} (\bibinfo {month}
  {Feb.}\ \bibinfo {year} {2013}),\
  \Eprint{http://arxiv.org/abs/1211.7148}{arXiv:1211.7148
  [astro-ph.HE]}\BibitemShut{NoStop}%
\bibitem{2013ApJ...765...56W}%
  \BibitemOpen
  \bibfield{author}{%
  \bibinfo {author} {\bibfnamefont{M.~C.}\ \bibnamefont{{Weisskopf}}}, \bibinfo
  {author} {\bibfnamefont{A.~F.}\ \bibnamefont{{Tennant}}}, \bibinfo {author}
  {\bibfnamefont{J.}~\bibnamefont{{Arons}}}, \bibinfo {author}
  {\bibfnamefont{R.}~\bibnamefont{{Blandford}}}, \bibinfo {author}
  {\bibfnamefont{R.}~\bibnamefont{{Buehler}}}, \bibinfo {author}
  {\bibfnamefont{P.}~\bibnamefont{{Caraveo}}}, \bibinfo {author}
  {\bibfnamefont{C.~C.}\ \bibnamefont{{Cheung}}}, \bibinfo {author}
  {\bibfnamefont{E.}~\bibnamefont{{Costa}}}, \bibinfo {author}
  {\bibfnamefont{A.}~\bibnamefont{{de Luca}}}, \bibinfo {author}
  {\bibfnamefont{C.}~\bibnamefont{{Ferrigno}}}, \bibinfo {author}
  {\bibfnamefont{H.}~\bibnamefont{{Fu}}}, \bibinfo {author}
  {\bibfnamefont{S.}~\bibnamefont{{Funk}}}, \bibinfo {author}
  {\bibfnamefont{M.}~\bibnamefont{{Habermehl}}}, \bibinfo {author}
  {\bibfnamefont{D.}~\bibnamefont{{Horns}}}, \bibinfo {author}
  {\bibfnamefont{J.~D.}\ \bibnamefont{{Linford}}}, \bibinfo {author}
  {\bibfnamefont{A.}~\bibnamefont{{Lobanov}}}, \bibinfo {author}
  {\bibfnamefont{C.}~\bibnamefont{{Max}}}, \bibinfo {author}
  {\bibfnamefont{R.}~\bibnamefont{{Mignani}}}, \bibinfo {author}
  {\bibfnamefont{S.~L.}\ \bibnamefont{{O'Dell}}}, \bibinfo {author}
  {\bibfnamefont{R.~W.}\ \bibnamefont{{Romani}}}, \bibinfo {author}
  {\bibfnamefont{E.}~\bibnamefont{{Striani}}}, \bibinfo {author}
  {\bibfnamefont{M.}~\bibnamefont{{Tavani}}}, \bibinfo {author}
  {\bibfnamefont{G.~B.}\ \bibnamefont{{Taylor}}}, \bibinfo {author}
  {\bibfnamefont{Y.}~\bibnamefont{{Uchiyama}}},\ and\ \bibinfo {author}
  {\bibfnamefont{Y.}~\bibnamefont{{Yuan}}},\ }%
  \bibfield{title}{%
  \enquote{\bibinfo {title} {{Chandra, Keck, and VLA Observations of the Crab
  Nebula during the 2011-April Gamma-Ray Flare}},}\ }%
  \bibfield{journal}{%
  \Doi{10.1088/0004-637X/765/1/56}{\bibinfo {journal} {ApJ}}\ }%
  \textbf{\bibinfo {volume} {765}},\ \bibinfo {eid} {56} (\bibinfo {month}
  {Mar.}\ \bibinfo {year} {2013}),\
  \Eprint{http://arxiv.org/abs/1211.3997}{arXiv:1211.3997
  [astro-ph.HE]}\BibitemShut{NoStop}%
\bibitem{1983MNRAS.205..593G}%
  \BibitemOpen
  \bibfield{author}{%
  \bibinfo {author} {\bibfnamefont{P.~W.}\ \bibnamefont{{Guilbert}}}, \bibinfo
  {author} {\bibfnamefont{A.~C.}\ \bibnamefont{{Fabian}}},\ and\ \bibinfo
  {author} {\bibfnamefont{M.~J.}\ \bibnamefont{{Rees}}},\ }%
  \bibfield{title}{%
  \enquote{\bibinfo {title} {{Spectral and variability constraints on compact
  sources}},}\ }%
  \bibfield{journal}{%
  \bibinfo {journal} {MNRAS}\ }%
  \textbf{\bibinfo {volume} {205}},\ \bibinfo {pages} {593--603} (\bibinfo
  {month} {Nov.}\ \bibinfo {year} {1983})\BibitemShut{NoStop}%
\bibitem{1996ApJ...457..253D}%
  \BibitemOpen
  \bibfield{author}{%
  \bibinfo {author} {\bibfnamefont{O.~C.}\ \bibnamefont{{de Jager}}}, \bibinfo
  {author} {\bibfnamefont{A.~K.}\ \bibnamefont{{Harding}}}, \bibinfo {author}
  {\bibfnamefont{P.~F.}\ \bibnamefont{{Michelson}}}, \bibinfo {author}
  {\bibfnamefont{H.~I.}\ \bibnamefont{{Nel}}}, \bibinfo {author}
  {\bibfnamefont{P.~L.}\ \bibnamefont{{Nolan}}}, \bibinfo {author}
  {\bibfnamefont{P.}~\bibnamefont{{Sreekumar}}},\ and\ \bibinfo {author}
  {\bibfnamefont{D.~J.}\ \bibnamefont{{Thompson}}},\ }%
  \bibfield{title}{%
  \enquote{\bibinfo {title} {{Gamma-Ray Observations of the Crab Nebula: A
  Study of the Synchro-Compton Spectrum}},}\ }%
  \bibfield{journal}{%
  \Doi{10.1086/176726}{\bibinfo {journal} {ApJ}}\ }%
  \textbf{\bibinfo {volume} {457}},\ \bibinfo {pages} {253} (\bibinfo {month}
  {Jan.}\ \bibinfo {year} {1996})\BibitemShut{NoStop}%
\bibitem{1970RvMP...42..237B}%
  \BibitemOpen
  \bibfield{author}{%
  \bibinfo {author} {\bibfnamefont{G.~R.}\ \bibnamefont{{Blumenthal}}}\ and\
  \bibinfo {author} {\bibfnamefont{R.~J.}\ \bibnamefont{{Gould}}},\ }%
  \bibfield{title}{%
  \enquote{\bibinfo {title} {{Bremsstrahlung, Synchrotron Radiation, and
  Compton Scattering of High-Energy Electrons Traversing Dilute Gases}},}\ }%
  \bibfield{journal}{%
  \Doi{10.1103/RevModPhys.42.237}{\bibinfo {journal} {Reviews of Modern
  Physics}}\ }%
  \textbf{\bibinfo {volume} {42}},\ \bibinfo {pages} {237--271} (\bibinfo
  {year} {1970})\BibitemShut{NoStop}%
\bibitem{2011MNRAS.414.2229B}%
  \BibitemOpen
  \bibfield{author}{%
  \bibinfo {author} {\bibfnamefont{W.}~\bibnamefont{{Bednarek}}}\ and\ \bibinfo
  {author} {\bibfnamefont{W.}~\bibnamefont{{Idec}}},\ }%
  \bibfield{title}{%
  \enquote{\bibinfo {title} {{On the variability of the GeV and multi-TeV
  gamma-ray emission from the Crab nebula}},}\ }%
  \bibfield{journal}{%
  \Doi{10.1111/j.1365-2966.2011.18539.x}{\bibinfo {journal} {MNRAS}}\ }%
  \textbf{\bibinfo {volume} {414}},\ \bibinfo {pages} {2229--2234} (\bibinfo
  {month} {Jul.}\ \bibinfo {year} {2011}),\
  \Eprint{http://arxiv.org/abs/1011.4176}{arXiv:1011.4176
  [astro-ph.HE]}\BibitemShut{NoStop}%
\bibitem{2011MNRAS.414.2017K}%
  \BibitemOpen
  \bibfield{author}{%
  \bibinfo {author} {\bibfnamefont{S.~S.}\ \bibnamefont{{Komissarov}}}\ and\
  \bibinfo {author} {\bibfnamefont{M.}~\bibnamefont{{Lyutikov}}},\ }%
  \bibfield{title}{%
  \enquote{\bibinfo {title} {{On the origin of variable gamma-ray emission from
  the Crab nebula}},}\ }%
  \bibfield{journal}{%
  \Doi{10.1111/j.1365-2966.2011.18516.x}{\bibinfo {journal} {MNRAS}}\ }%
  \textbf{\bibinfo {volume} {414}},\ \bibinfo {pages} {2017--2028} (\bibinfo
  {month} {Jul.}\ \bibinfo {year} {2011}),\
  \Eprint{http://arxiv.org/abs/1011.1800}{arXiv:1011.1800
  [astro-ph.HE]}\BibitemShut{NoStop}%
\bibitem{2011ApJ...730L..15Y}%
  \BibitemOpen
  \bibfield{author}{%
  \bibinfo {author} {\bibfnamefont{Q.}~\bibnamefont{{Yuan}}}, \bibinfo {author}
  {\bibfnamefont{P.-F.}\ \bibnamefont{{Yin}}}, \bibinfo {author}
  {\bibfnamefont{X.-F.}\ \bibnamefont{{Wu}}}, \bibinfo {author}
  {\bibfnamefont{X.-J.}\ \bibnamefont{{Bi}}}, \bibinfo {author}
  {\bibfnamefont{S.}~\bibnamefont{{Liu}}},\ and\ \bibinfo {author}
  {\bibfnamefont{B.}~\bibnamefont{{Zhang}}},\ }%
  \bibfield{title}{%
  \enquote{\bibinfo {title} {{A Statistical Model for the {$\gamma$}-ray
  Variability of the Crab Nebula}},}\ }%
  \bibfield{journal}{%
  \Doi{10.1088/2041-8205/730/2/L15}{\bibinfo {journal} {ApJL}}\ }%
  \textbf{\bibinfo {volume} {730}},\ \bibinfo {eid} {L15} (\bibinfo {month}
  {Apr.}\ \bibinfo {year} {2011}),\
  \Eprint{http://arxiv.org/abs/1012.1395}{arXiv:1012.1395
  [astro-ph.HE]}\BibitemShut{NoStop}%
\bibitem{2012MNRAS.422.3118L}%
  \BibitemOpen
  \bibfield{author}{%
  \bibinfo {author} {\bibfnamefont{M.}~\bibnamefont{{Lyutikov}}}, \bibinfo
  {author} {\bibfnamefont{D.}~\bibnamefont{{Balsara}}},\ and\ \bibinfo {author}
  {\bibfnamefont{C.}~\bibnamefont{{Matthews}}},\ }%
  \bibfield{title}{%
  \enquote{\bibinfo {title} {{Crab GeV flares from the corrugated termination
  shock}},}\ }%
  \bibfield{journal}{%
  \Doi{10.1111/j.1365-2966.2012.20831.x}{\bibinfo {journal} {MNRAS}}\ }%
  \textbf{\bibinfo {volume} {422}},\ \bibinfo {pages} {3118--3129} (\bibinfo
  {month} {Jun.}\ \bibinfo {year} {2012}),\
  \Eprint{http://arxiv.org/abs/1109.1204}{arXiv:1109.1204
  [astro-ph.HE]}\BibitemShut{NoStop}%
\bibitem{2012MNRAS.426.1374C}%
  \BibitemOpen
  \bibfield{author}{%
  \bibinfo {author} {\bibfnamefont{E.}~\bibnamefont{{Clausen-Brown}}}\ and\
  \bibinfo {author} {\bibfnamefont{M.}~\bibnamefont{{Lyutikov}}},\ }%
  \bibfield{title}{%
  \enquote{\bibinfo {title} {{Crab nebula gamma-ray flares as relativistic
  reconnection minijets}},}\ }%
  \bibfield{journal}{%
  \Doi{10.1111/j.1365-2966.2012.21349.x}{\bibinfo {journal} {MNRAS}}\ }%
  \textbf{\bibinfo {volume} {426}},\ \bibinfo {pages} {1374--1384} (\bibinfo
  {month} {Oct.}\ \bibinfo {year} {2012}),\
  \Eprint{http://arxiv.org/abs/1205.5094}{arXiv:1205.5094
  [astro-ph.HE]}\BibitemShut{NoStop}%
\bibitem{2012MNRAS.427.1497L}%
  \BibitemOpen
  \bibfield{author}{%
  \bibinfo {author} {\bibfnamefont{Y.~E.}\ \bibnamefont{{Lyubarsky}}},\ }%
  \bibfield{title}{%
  \enquote{\bibinfo {title} {{Highly magnetized region in pulsar wind nebulae
  and origin of the Crab gamma-ray flares}},}\ }%
  \bibfield{journal}{%
  \Doi{10.1111/j.1365-2966.2012.22097.x}{\bibinfo {journal} {MNRAS}}\ }%
  \textbf{\bibinfo {volume} {427}},\ \bibinfo {pages} {1497--1502} (\bibinfo
  {month} {Dec.}\ \bibinfo {year} {2012}),\
  \Eprint{http://arxiv.org/abs/1209.1589}{arXiv:1209.1589
  [astro-ph.HE]}\BibitemShut{NoStop}%
\bibitem{2002ApJ...577L..49H}%
  \BibitemOpen
  \bibfield{author}{%
  \bibinfo {author} {\bibfnamefont{J.~J.}\ \bibnamefont{{Hester}}}, \bibinfo
  {author} {\bibfnamefont{K.}~\bibnamefont{{Mori}}}, \bibinfo {author}
  {\bibfnamefont{D.}~\bibnamefont{{Burrows}}}, \bibinfo {author}
  {\bibfnamefont{J.~S.}\ \bibnamefont{{Gallagher}}}, \bibinfo {author}
  {\bibfnamefont{J.~R.}\ \bibnamefont{{Graham}}}, \bibinfo {author}
  {\bibfnamefont{M.}~\bibnamefont{{Halverson}}}, \bibinfo {author}
  {\bibfnamefont{A.}~\bibnamefont{{Kader}}}, \bibinfo {author}
  {\bibfnamefont{F.~C.}\ \bibnamefont{{Michel}}},\ and\ \bibinfo {author}
  {\bibfnamefont{P.}~\bibnamefont{{Scowen}}},\ }%
  \bibfield{title}{%
  \enquote{\bibinfo {title} {{Hubble Space Telescope and Chandra Monitoring of
  the Crab Synchrotron Nebula}},}\ }%
  \bibfield{journal}{%
  \Doi{10.1086/344132}{\bibinfo {journal} {ApJL}}\ }%
  \textbf{\bibinfo {volume} {577}},\ \bibinfo {pages} {L49--L52} (\bibinfo
  {month} {Sep.}\ \bibinfo {year} {2002})\BibitemShut{NoStop}%
\bibitem{2011ApJ...737L..40U}%
  \BibitemOpen
  \bibfield{author}{%
  \bibinfo {author} {\bibfnamefont{D.~A.}\ \bibnamefont{{Uzdensky}}}, \bibinfo
  {author} {\bibfnamefont{B.}~\bibnamefont{{Cerutti}}},\ and\ \bibinfo {author}
  {\bibfnamefont{M.~C.}\ \bibnamefont{{Begelman}}},\ }%
  \bibfield{title}{%
  \enquote{\bibinfo {title} {{Reconnection-powered Linear Accelerator and
  Gamma-Ray Flares in the Crab Nebula}},}\ }%
  \bibfield{journal}{%
  \Doi{10.1088/2041-8205/737/2/L40}{\bibinfo {journal} {ApJL}}\ }%
  \textbf{\bibinfo {volume} {737}},\ \bibinfo {eid} {L40} (\bibinfo {month}
  {Aug.}\ \bibinfo {year} {2011}),\
  \Eprint{http://arxiv.org/abs/1105.0942}{arXiv:1105.0942
  [astro-ph.HE]}\BibitemShut{NoStop}%
\bibitem{2012ApJ...746..148C}%
  \BibitemOpen
  \bibfield{author}{%
  \bibinfo {author} {\bibfnamefont{B.}~\bibnamefont{{Cerutti}}}, \bibinfo
  {author} {\bibfnamefont{D.~A.}\ \bibnamefont{{Uzdensky}}},\ and\ \bibinfo
  {author} {\bibfnamefont{M.~C.}\ \bibnamefont{{Begelman}}},\ }%
  \bibfield{title}{%
  \enquote{\bibinfo {title} {{Extreme Particle Acceleration in Magnetic
  Reconnection Layers: Application to the Gamma-Ray Flares in the Crab
  Nebula}},}\ }%
  \bibfield{journal}{%
  \Doi{10.1088/0004-637X/746/2/148}{\bibinfo {journal} {ApJ}}\ }%
  \textbf{\bibinfo {volume} {746}},\ \bibinfo {eid} {148} (\bibinfo {month}
  {Feb.}\ \bibinfo {year} {2012}),\
  \Eprint{http://arxiv.org/abs/1110.0557}{arXiv:1110.0557
  [astro-ph.HE]}\BibitemShut{NoStop}%
\bibitem{2013ApJ...770..147C}%
  \BibitemOpen
  \bibfield{author}{%
  \bibinfo {author} {\bibfnamefont{B.}~\bibnamefont{{Cerutti}}}, \bibinfo
  {author} {\bibfnamefont{G.~R.}\ \bibnamefont{{Werner}}}, \bibinfo {author}
  {\bibfnamefont{D.~A.}\ \bibnamefont{{Uzdensky}}},\ and\ \bibinfo {author}
  {\bibfnamefont{M.~C.}\ \bibnamefont{{Begelman}}},\ }%
  \bibfield{title}{%
  \enquote{\bibinfo {title} {{Simulations of Particle Acceleration beyond the
  Classical Synchrotron Burnoff Limit in Magnetic Reconnection: An Explanation
  of the Crab Flares}},}\ }%
  \bibfield{journal}{%
  \Doi{10.1088/0004-637X/770/2/147}{\bibinfo {journal} {ApJ}}\ }%
  \textbf{\bibinfo {volume} {770}},\ \bibinfo {eid} {147} (\bibinfo {month}
  {Jun.}\ \bibinfo {year} {2013}),\
  \Eprint{http://arxiv.org/abs/1302.6247}{arXiv:1302.6247
  [astro-ph.HE]}\BibitemShut{NoStop}%
\bibitem{2013arXiv1311.2605C}%
  \BibitemOpen
  \bibfield{author}{%
  \bibinfo {author} {\bibfnamefont{B.}~\bibnamefont{{Cerutti}}}, \bibinfo
  {author} {\bibfnamefont{G.~R.}\ \bibnamefont{{Werner}}}, \bibinfo {author}
  {\bibfnamefont{D.~A.}\ \bibnamefont{{Uzdensky}}},\ and\ \bibinfo {author}
  {\bibfnamefont{M.~C.}\ \bibnamefont{{Begelman}}},\ }%
  \bibfield{title}{%
  \enquote{\bibinfo {title} {{Three-dimensional relativistic pair plasma
  reconnection with radiative feedback in the Crab Nebula}},}\ }%
  \bibfield{journal}{%
  \bibinfo {journal} {Accepted in ApJ}}%
   (\bibinfo {month} {Nov.}\ \bibinfo {year} {2013}),\
  \Eprint{http://arxiv.org/abs/1311.2605}{arXiv:1311.2605
  [astro-ph.HE]}\BibitemShut{NoStop}%
\bibitem{2004PhRvL..92r1101K}%
  \BibitemOpen
  \bibfield{author}{%
  \bibinfo {author} {\bibfnamefont{J.~G.}\ \bibnamefont{{Kirk}}},\ }%
  \bibfield{title}{%
  \enquote{\bibinfo {title} {{Particle Acceleration in Relativistic Current
  Sheets}},}\ }%
  \bibfield{journal}{%
  \Doi{10.1103/PhysRevLett.92.181101}{\bibinfo {journal} {Phys. Rev. Lett.}}\
  }%
  \textbf{\bibinfo {volume} {92}},\ \bibinfo {eid} {181101} (\bibinfo {month}
  {May}\ \bibinfo {year} {2004}),\
  \Eprint{http://arxiv.org/abs/astro-ph/0403516}{astro-ph/0403516}\BibitemShut%
{NoStop}%
\bibitem{2007A&A...472..219C}%
  \BibitemOpen
  \bibfield{author}{%
  \bibinfo {author} {\bibfnamefont{I.}~\bibnamefont{{Contopoulos}}},\ }%
  \bibfield{title}{%
  \enquote{\bibinfo {title} {{The magnetic field topology in the reconnecting
  pulsar magnetosphere}},}\ }%
  \bibfield{journal}{%
  \Doi{10.1051/0004-6361:20077167}{\bibinfo {journal} {A\&A}}\ }%
  \textbf{\bibinfo {volume} {472}},\ \bibinfo {pages} {219--223} (\bibinfo
  {month} {Sep.}\ \bibinfo {year} {2007}),\
  \Eprint{http://arxiv.org/abs/0706.2255}{arXiv:0706.2255}\BibitemShut{NoStop}%
\bibitem{2009PhRvL.103g5002J}%
  \BibitemOpen
  \bibfield{author}{%
  \bibinfo {author} {\bibfnamefont{C.~H.}\ \bibnamefont{{Jaroschek}}}\ and\
  \bibinfo {author} {\bibfnamefont{M.}~\bibnamefont{{Hoshino}}},\ }%
  \bibfield{title}{%
  \enquote{\bibinfo {title} {{Radiation-Dominated Relativistic Current
  Sheets}},}\ }%
  \bibfield{journal}{%
  \Doi{10.1103/PhysRevLett.103.075002}{\bibinfo {journal} {Physical Review
  Letters}}\ }%
  \textbf{\bibinfo {volume} {103}},\ \bibinfo {eid} {075002} (\bibinfo {month}
  {Aug.}\ \bibinfo {year} {2009})\BibitemShut{NoStop}%
\bibitem{2010NJPh...12l3005T}%
  \BibitemOpen
  \bibfield{author}{%
  \bibinfo {author} {\bibfnamefont{M.}~\bibnamefont{{Tamburini}}}, \bibinfo
  {author} {\bibfnamefont{F.}~\bibnamefont{{Pegoraro}}}, \bibinfo {author}
  {\bibfnamefont{A.}~\bibnamefont{{Di Piazza}}}, \bibinfo {author}
  {\bibfnamefont{C.~H.}\ \bibnamefont{{Keitel}}},\ and\ \bibinfo {author}
  {\bibfnamefont{A.}~\bibnamefont{{Macchi}}},\ }%
  \bibfield{title}{%
  \enquote{\bibinfo {title} {{Radiation reaction effects on radiation pressure
  acceleration}},}\ }%
  \bibfield{journal}{%
  \Doi{10.1088/1367-2630/12/12/123005}{\bibinfo {journal} {New Journal of
  Physics}}\ }%
  \textbf{\bibinfo {volume} {12}},\ \bibinfo {eid} {123005} (\bibinfo {month}
  {Dec.}\ \bibinfo {year} {2010}),\
  \Eprint{http://arxiv.org/abs/1008.1685}{arXiv:1008.1685
  [physics.plasm-ph]}\BibitemShut{NoStop}%
\bibitem{1975ctf..book.....L}%
  \BibitemOpen
  \bibfield{author}{%
  \bibinfo {author} {\bibfnamefont{L.~D.}\ \bibnamefont{{Landau}}}\ and\
  \bibinfo {author} {\bibfnamefont{E.~M.}\ \bibnamefont{{Lifshitz}}},\ }%
  \emph{\bibinfo {title} {Course of theoretical physics - Pergamon
  International Library of Science, Technology, Engineering and Social Studies,
  Oxford: Pergamon Press, 1975, 4th rev.engl.ed.}}\ (\bibinfo {year}
  {1975})\BibitemShut{NoStop}%
\bibitem{1966ITAP...14..302Y}%
  \BibitemOpen
  \bibfield{author}{%
  \bibinfo {author} {\bibfnamefont{K.}~\bibnamefont{{Yee}}},\ }%
  \bibfield{title}{%
  \enquote{\bibinfo {title} {{Numerical solution of inital boundary value
  problems involving maxwell's equations in isotropic media}},}\ }%
  \bibfield{journal}{%
  \Doi{10.1109/TAP.1966.1138693}{\bibinfo {journal} {IEEE Transactions on
  Antennas and Propagation}}\ }%
  \textbf{\bibinfo {volume} {14}},\ \bibinfo {pages} {302--307} (\bibinfo
  {month} {May}\ \bibinfo {year} {1966})\BibitemShut{NoStop}%
\bibitem{2003ApJ...591..366K}%
  \BibitemOpen
  \bibfield{author}{%
  \bibinfo {author} {\bibfnamefont{J.~G.}\ \bibnamefont{{Kirk}}}\ and\ \bibinfo
  {author} {\bibfnamefont{O.}~\bibnamefont{{Skj{\ae}raasen}}},\ }%
  \bibfield{title}{%
  \enquote{\bibinfo {title} {{Dissipation in Poynting-Flux-dominated Flows: The
  {$\sigma$}-Problem of the Crab Pulsar Wind}},}\ }%
  \bibfield{journal}{%
  \Doi{10.1086/375215}{\bibinfo {journal} {ApJ}}\ }%
  \textbf{\bibinfo {volume} {591}},\ \bibinfo {pages} {366--379} (\bibinfo
  {month} {Jul.}\ \bibinfo {year} {2003}),\
  \Eprint{http://arxiv.org/abs/astro-ph/0303194}{astro-ph/0303194}\BibitemShut%
{NoStop}%
\bibitem{2008ApJ...677..530Z}%
  \BibitemOpen
  \bibfield{author}{%
  \bibinfo {author} {\bibfnamefont{S.}~\bibnamefont{{Zenitani}}}\ and\ \bibinfo
  {author} {\bibfnamefont{M.}~\bibnamefont{{Hoshino}}},\ }%
  \bibfield{title}{%
  \enquote{\bibinfo {title} {{The Role of the Guide Field in Relativistic Pair
  Plasma Reconnection}},}\ }%
  \bibfield{journal}{%
  \Doi{10.1086/528708}{\bibinfo {journal} {ApJ}}\ }%
  \textbf{\bibinfo {volume} {677}},\ \bibinfo {pages} {530--544} (\bibinfo
  {month} {Apr.}\ \bibinfo {year} {2008}),\
  \Eprint{http://arxiv.org/abs/0712.2016}{arXiv:0712.2016}\BibitemShut{NoStop}%
\bibitem{1965JGR....70.4219S}%
  \BibitemOpen
  \bibfield{author}{%
  \bibinfo {author} {\bibfnamefont{T.~W.}\ \bibnamefont{{Speiser}}},\ }%
  \bibfield{title}{%
  \enquote{\bibinfo {title} {{Particle Trajectories in Model Current Sheets, 1,
  Analytical Solutions}},}\ }%
  \bibfield{journal}{%
  \Doi{10.1029/JZ070i017p04219}{\bibinfo {journal} {J. Geophys. Res.}}\ }%
  \textbf{\bibinfo {volume} {70}},\ \bibinfo {pages} {4219--4226} (\bibinfo
  {month} {Sep.}\ \bibinfo {year} {1965})\BibitemShut{NoStop}%
\bibitem{2012ApJ...754L..33C}%
  \BibitemOpen
  \bibfield{author}{%
  \bibinfo {author} {\bibfnamefont{B.}~\bibnamefont{{Cerutti}}}, \bibinfo
  {author} {\bibfnamefont{G.~R.}\ \bibnamefont{{Werner}}}, \bibinfo {author}
  {\bibfnamefont{D.~A.}\ \bibnamefont{{Uzdensky}}},\ and\ \bibinfo {author}
  {\bibfnamefont{M.~C.}\ \bibnamefont{{Begelman}}},\ }%
  \bibfield{title}{%
  \enquote{\bibinfo {title} {{Beaming and Rapid Variability of High-energy
  Radiation from Relativistic Pair Plasma Reconnection}},}\ }%
  \bibfield{journal}{%
  \Doi{10.1088/2041-8205/754/2/L33}{\bibinfo {journal} {ApJL}}\ }%
  \textbf{\bibinfo {volume} {754}},\ \bibinfo {eid} {L33} (\bibinfo {month}
  {Aug.}\ \bibinfo {year} {2012}),\
  \Eprint{http://arxiv.org/abs/1205.3210}{arXiv:1205.3210
  [astro-ph.HE]}\BibitemShut{NoStop}%
\bibitem{2012MNRAS.425.2519N}%
  \BibitemOpen
  \bibfield{author}{%
  \bibinfo {author} {\bibfnamefont{K.}~\bibnamefont{{Nalewajko}}}, \bibinfo
  {author} {\bibfnamefont{M.~C.}\ \bibnamefont{{Begelman}}}, \bibinfo {author}
  {\bibfnamefont{B.}~\bibnamefont{{Cerutti}}}, \bibinfo {author}
  {\bibfnamefont{D.~A.}\ \bibnamefont{{Uzdensky}}},\ and\ \bibinfo {author}
  {\bibfnamefont{M.}~\bibnamefont{{Sikora}}},\ }%
  \bibfield{title}{%
  \enquote{\bibinfo {title} {{Energetic constraints on a rapid gamma-ray flare
  in PKS 1222+216}},}\ }%
  \bibfield{journal}{%
  \Doi{10.1111/j.1365-2966.2012.21721.x}{\bibinfo {journal} {MNRAS}}\ }%
  \textbf{\bibinfo {volume} {425}},\ \bibinfo {pages} {2519--2529} (\bibinfo
  {month} {Oct.}\ \bibinfo {year} {2012}),\
  \Eprint{http://arxiv.org/abs/1202.2123}{arXiv:1202.2123
  [astro-ph.HE]}\BibitemShut{NoStop}%
\bibitem{2013MNRAS.431L..48P}%
  \BibitemOpen
  \bibfield{author}{%
  \bibinfo {author} {\bibfnamefont{O.}~\bibnamefont{{Porth}}}, \bibinfo
  {author} {\bibfnamefont{S.~S.}\ \bibnamefont{{Komissarov}}},\ and\ \bibinfo
  {author} {\bibfnamefont{R.}~\bibnamefont{{Keppens}}},\ }%
  \bibfield{title}{%
  \enquote{\bibinfo {title} {{Solution to the sigma problem of pulsar wind
  nebulae}},}\ }%
  \bibfield{journal}{%
  \Doi{10.1093/mnrasl/slt006}{\bibinfo {journal} {MNRAS}}\ }%
  \textbf{\bibinfo {volume} {431}},\ \bibinfo {pages} {L48--L52} (\bibinfo
  {month} {Apr.}\ \bibinfo {year} {2013}),\
  \Eprint{http://arxiv.org/abs/1212.1382}{arXiv:1212.1382
  [astro-ph.HE]}\BibitemShut{NoStop}%
\bibitem{2013arXiv1310.2531P}%
  \BibitemOpen
  \bibfield{author}{%
  \bibinfo {author} {\bibfnamefont{O.}~\bibnamefont{{Porth}}}, \bibinfo
  {author} {\bibfnamefont{S.~S.}\ \bibnamefont{{Komissarov}}},\ and\ \bibinfo
  {author} {\bibfnamefont{R.}~\bibnamefont{{Keppens}}},\ }%
  \bibfield{title}{%
  \enquote{\bibinfo {title} {{Three-Dimensional Magnetohydrodynamic Simulations
  of the Crab Nebula}},}\ }%
  \bibfield{journal}{%
  \bibinfo {journal} {ArXiv e-prints}}%
   (\bibinfo {month} {Oct.}\ \bibinfo {year} {2013}),\
  \Eprint{http://arxiv.org/abs/1310.2531}{arXiv:1310.2531
  [astro-ph.HE]}\BibitemShut{NoStop}%
\bibitem{1998ApJ...493..291B}%
  \BibitemOpen
  \bibfield{author}{%
  \bibinfo {author} {\bibfnamefont{M.~C.}\ \bibnamefont{{Begelman}}},\ }%
  \bibfield{title}{%
  \enquote{\bibinfo {title} {{Instability of Toroidal Magnetic Field in Jets
  and Plerions}},}\ }%
  \bibfield{journal}{%
  \Doi{10.1086/305119}{\bibinfo {journal} {ApJ}}\ }%
  \textbf{\bibinfo {volume} {493}},\ \bibinfo {pages} {291} (\bibinfo {month}
  {Jan.}\ \bibinfo {year} {1998}),\
  \Eprint{http://arxiv.org/abs/astro-ph/9708142}{astro-ph/9708142}\BibitemShut%
{NoStop}%
\end{thebibliography}%

\end{document}